\input psfig
\input mn.tex


\def\d{{\rm d}}
\def\bandg{{\beta + \gamma}}

\def\fr#1#2{\textstyle {#1\over #2}\displaystyle}

\def\piby2{{\pi \over 2}}

%
%
\def\spose#1{\hbox to 0pt{#1\hss}}
\def\lta{\mathrel{\spose{\lower 3pt\hbox{$\sim$}}
    \raise 2.0pt\hbox{$<$}}}
\def\gta{\mathrel{\spose{\lower 3pt\hbox{$\sim$}}
    \raise 2.0pt\hbox{$>$}}}
\def\today{\ifcase\month\or
 January\or February\or March\or April\or May\or June\or
 July\or August\or September\or October\or November\or December\fi
 \space\number\day, \number\year}
\newdimen\hssize
\hssize=8.4truecm  
\newdimen\hdsize
\hdsize=16.8truecm    


\newcount\eqnumber
\eqnumber=1
\def\chaphead{}
 
\def\new{\hbox{(\rm\chaphead\the\eqnumber)}\global\advance\eqnumber by 1}
 
\def\first{\hbox{(\rm\chaphead\the\eqnumber a)}\global\advance\eqnumber by 1}
\def\last#1{\advance\eqnumber by -1 \hbox{(\rm\chaphead\the\eqnumber#1)}\advance
     \eqnumber by 1}
 
\def\ref#1{\advance\eqnumber by -#1 \chaphead\the\eqnumber
     \advance\eqnumber by #1}
 
\def\nref#1{\advance\eqnumber by -#1 \chaphead\the\eqnumber
     \advance\eqnumber by #1}

\def\eqnam#1{\xdef#1{\chaphead\the\eqnumber}}
 
 

\pageoffset{-0.85truecm}{-1.05truecm}



\pagerange{}
\pubyear{1996}
\volume{000, 000--000}


\begintopmatter

\title{Simple Three-Integral Scale-Free Galaxy Models}

\author{N.W.\ Evans,$^1$ R.M.\ H\"afner$^1$ and P.T.\ de Zeeuw$^2$}

\vskip0.15truecm
\affiliation{$^1$Theoretical Physics, Department of Physics, 1 Keble Road,
                 Oxford, OX1 3NP}
\vskip0.15truecm
\affiliation{$^2$Sterrewacht Leiden, Postbus 9513, 2300 RA Leiden,
                 The Netherlands}

\shortauthor{N.W.\ Evans, R.M.\ H\"afner and P.T.\ de Zeeuw} 

\shorttitle{Simple Three-Integral Scale-Free Galaxy Models} 



\abstract{The Jeans equations give the second moments or stresses
required to support a stellar population against the gravity field.  A
general solution of the Jeans equations for arbitrary axisymmetric
scale-free densities in flattened scale-free potentials is given.  A
two-parameter subset of the solution for the second moments for the
self-consistent density of the power-law models, which have exactly
spheroidal equipotentials, is examined in detail. In the spherical
limit, the potential of these models reduces to that of the singular
power-law spheres. We build the physical three-integral distribution
functions that correspond to the flattened stellar components.

Next, we attack the problem of finding distribution functions
associated with the Jeans solutions in flattened scale-free
potentials.  The third or partial integral introduced by de Zeeuw,
Evans \& Schwarzschild for Binney's model is generalised to thin and
near-thin orbits moving in arbitrary axisymmetric scale-free
potentials.  The partial integral is a modification of the total
angular momentum.  For the self-consistent power-law models, we show
how this enables the construction of simple three-integral
distribution functions. The connexion between these approximate
distribution functions and the Jeans solutions is discussed in some
detail.}


\maketitle  

\section{1 Introduction}

Evidence both from hydrodynamical modelling of ellipticals (e.g.,
Binney, Davies \& Illingworth 1990; van der Marel 1991) and the
kinematics of the stellar halo of the Milky Way (e.g., Norris 1986;
Morrison, Flynn \& Freeman 1990) indicates that the phase space
distribution functions (DFs) of elliptical galaxies and the haloes of
spiral galaxies depend on three isolating integrals of motion.

General flattened potentials support fewer than three global isolating
integrals. In triaxial systems, only the energy $E$ is globally
conserved. In axisymmetric potentials, the angular momentum component
parallel to the symmetry axis $L_z$ is also an invariant. Nonetheless,
most stars do admit a third integral of motion that is a
generalisation of the total angular momentum $L$ (c.f., Saaf 1968;
Innanen \& Papp 1977; Lupton \& Gunn 1987). The existence of chaotic
trajectories prevents the extension of an analytic third integral to
all the orbits. So, building three-integral DFs is often difficult in
axisymmetric potentials. The problem is still more severe for triaxial
systems (Schwarzschild 1979).

The natural shape of ellipticals and the haloes of spiral galaxies may
well be triaxial. Nonetheless, it is still both useful and
realistic to consider the somewhat simpler problem of how to construct
three-integral axisymmetric stellar systems.  Recently, de Zeeuw,
Evans \& Schwarzschild (1996, hereafter ZES) showed how to find a
third, partial, integral of good accuracy for thin and near-thin tubes
for oblate scale-free potentials with flat rotation curves. They used
the partial integral to build three-integral DFs and identified
a class of physical solutions to the Jeans equations. The aim of
this paper is to extend their methods and results to arbitrary
axisymmetric scale-free potentials, with particular emphasis on the
models with spheroidal equipotentials known as the power-law models
(Evans 1994, hereafter E94).

The paper is arranged as follows. In Section 2, the Jeans approach is
applied to yield the complete solution for the stresses or second
order velocity moments that can support flattened populations of stars
in scale-free potentials. A particularly simple two-parameter solution
is identified for the self-consistent power-law models. Section 3
investigates the spherical limit. New families of three-integral DFs
are presented and used to isolate the Jeans solutions of physical
interest. With the insight gained from this simpler problem, we begin
the harder job of constructing three-integral DFs for flattened
scale-free potentials in Section 4. The third or partial integral
introduced by ZES for the scale-free logarithmic potential (e.g.,
Binney \& Tremaine 1987; see also Richstone 1980) is generalised to
arbitrary axisymmetric scale-free potentials. We show by detailed
numerical orbit integrations that it is a good integral for the thin
and near-thin tube orbits, although it is not so well-conserved for
fat tube orbits. Three integral DFs are then built in Section 5 and
shown to correspond to a subset of the earlier Jeans solutions.

\eqnumber =1
\def\chaphead{\hbox{2.}}
\section{2 The Jeans Approach} 

In this section, we use standard spherical coordinates $(r,\theta,
\phi)$, with $\theta$ measured from the axis of symmetry and $\phi$
the azimuthal angle.

\subsection{2.1 Scale-free potentials and densities} 

Scale-free axisymmetric potentials have the form
\eqnam{\genpot}
$$\Phi = -{v_0^2 \over \beta r^\beta g^{\beta /2}(\theta)},\eqno\new$$
where $g(\theta)$ is an arbitrary function that describes the shape of
the equipotentials.  The circular velocity $v_{\rm circ}$ in the
equatorial plane varies like $r^{-\beta/2}$. So, models with $\beta
<0$ have rising rotation curves, whereas models with $\beta >0$ have
falling rotation curves. When $\beta =0$, the scale-free power-law
potential becomes logarithmic, and we recover the special case already
studied by ZES. The relevant range of $\beta$ is from $1$
corresponding to the outer Keplerian envelopes of star clusters to
$-2$ corresponding to the shallow cusps of boxy ellipticals.
Henceforth, we use units in which the velocity scale $v_0$ is unity.

It is often useful to consider tracer populations of stars moving in
an external gravity field rather than the self-consistent stellar
density generated by Poisson's equation. So, let us take the density to
have the general scale-free form:
\eqnam{\genden}
$$\rho = {h(\theta)\over r^\gamma g^{2 + \beta/2}(\theta)}.\eqno\new$$
Here, $\gamma$ is a constant, which prescribes the radial fall-off of
the density, while $h(\theta)$ is an arbitrary function. Both $g(\theta)$
and $h(\theta)$ are symmetric with respect to the 
equatorial plane, i.e., $g(\pi-\theta)=
g(\theta)$ and $h(\pi-\theta)=h(\theta)$.  When $\rho$ is the
self-consistent density, then $\gamma =2 + \beta$ and $h(\theta)$ 
and $g(\theta)$ are related by ($4\pi G = 1$)
$$\eqalign{
h(\theta) =& (1-\beta) g^2(\theta) +\fr12 g(\theta)g'(\theta) \cot\theta  \cr
          &+\fr12g(\theta)g''(\theta) -\fr12 g'(\theta) g'(\theta)(1 +
          \fr12 \beta) . \cr}
                                                       \eqno\new$$
When $\gamma\not=2 + \beta$, the density (\genden) is greater than the
density associated with the potential (\genpot) either at large radii
($\gamma<2 + \beta$) or at small radii ($\gamma>2 + \beta$). However,
we are often interested in the dynamics in particular r\'egimes -- such
as the outer reaches or the cusp -- and so this is not a grave
drawback, provided the difficulty occurs outside the r\'egime under
scrutiny.

One pleasing choice for the arbitrary function $g (\theta)$ is
\eqnam{\powergtheta}
$$g(\theta) = \sin^2\theta + {\cos^2\theta \over q^2},    \eqno\new$$
so that the equipotentials are similar concentric spheroids with axis
ratio $q$. These are recognised as the power-law models introduced in E94. 
The associated self-consistent density is of the form (\genden), with 
\eqnam{\powerhtheta}
$$h(\theta) = Q \bigl\{(1- \beta q^2) \sin^2\theta  + 
[2-Q(1+\beta)] \cos^2\theta \bigr\},\eqno\new$$
with $Q= q^{-2}$. In the spherical limit, $q = Q =1$ and so
$h(\theta) = 1- \beta$.

\subsection{2.2 General solution of the Jeans equations} 

The potential (\genpot) and density (\genden) have the felicitous
attribute of scale-freeness. Their properties at radius $r' = kr$
are just a magnification of those at radius $r$. Scale-free density 
distributions may have DFs that are not scale-free. For example,
scale-free spheres can possess DFs built according to the instructions
provided by Osipkov (1979) and Merritt (1985). These have an anisotropy
radius, at which the properties of the velocity dispersion tensor 
change. However, let us assume that the associated DFs are also 
scale-free (e.g., Richstone 1980; White 1985; ZES), then the stresses 
have the following form:
\eqnam{\stressansatz}
$$\eqalign{
\rho \langle v_r^2 \rangle &= {F_1(\theta) \over r^\bandg g^{2+\beta}
(\theta)},\cr
\rho \langle v_\theta^2 \rangle &= 
                {F_3(\theta)\over r^\bandg g^{2+\beta}(\theta)}, \cr}
\qquad
\eqalign{
\rho \langle v_r v_\theta \rangle 
   &= {F_2(\theta) \over r^\bandg g^{2+\beta}(\theta)}, \cr
\rho \langle v_\phi^2 \rangle &= {F_4(\theta) \over 
r^\bandg g^{2+\beta}(\theta)},\cr}                         \eqno\new$$
where $F_1$, $F_2$, $F_3$, and $F_4$ are functions of $\theta$,
fulfilling the following conditions
\eqnam{\symmetryconstraint}
$$\eqalign{
F_1(\pi\!-\!\theta) &= F_1(\theta), \cr
F_2(\pi\!-\!\theta) &= -F_2(\theta), \cr} 
\qquad\qquad
\eqalign{
F_3(\pi\!-\!\theta) &= F_3(\theta), \cr
F_4(\pi\!-\!\theta) &= F_4(\theta). \cr}                     \eqno\new$$
As discussed in ZES, these constraints guarantee that the stresses are
symmetric with respect to the equatorial plane. Substitution of the
Ansatz (\stressansatz) into the Jeans equations (ZES, equation 2.4)
reduces them to two coupled first-order ordinary differential
equations:
\eqnam{\scalejeans}
$$\eqalign{
&F'_2(\theta) 
 + \Bigl[ \cot \theta - {(2+\beta)g'(\theta) \over g(\theta)} \Bigr] 
 F_2(\theta) \cr
 &+ (2 - \beta - \gamma) F_1( \theta)  -F_3(\theta) - F_4(\theta) = 
-h(\theta), \cr}                 \eqno\first$$
$$\eqalign{
&F'_3(\theta) 
+ \Bigl[ \cot \theta - {(2+\beta)g'(\theta) \over g(\theta)} \Bigr] 
 F_3(\theta) \cr
&+(3- \beta -\gamma ) F_2(\theta) - F_4(\theta) \cot\theta  = 
  -{h(\theta) g'(\theta) \over 2g(\theta)}. \cr}          \eqno\last b$$
The four functions $F_1(\theta), \ldots, F_4(\theta)$ are therefore 
subject to the above two restrictions. We are at liberty to pick two of
the functions arbitrarily and solve (\scalejeans) for the other
two. We choose to prescribe $F_1(\theta)$ and $F_2(\theta)$.
Proceeding as in ZES, we use equation (\scalejeans a) to eliminate
$F_4(\theta)$ from (\scalejeans b). It is straightforward to integrate
the resulting first order differential equation for $F_3$ and then to
substitute the result in (\scalejeans a) to obtain $F_4$. We find:
\eqnam{\generalsolution}
$$F_3(\theta) = I(\theta) + J(\theta) + F_2(\theta) \cot\theta, \eqno\first$$
$$\eqalign{
F_4(\theta) &= h(\theta) - I(\theta) -J(\theta) \cr
& -(\beta + \gamma\!-\!2) F_1(\theta) + g^{\beta + 2}(\theta) 
{{\rm d}\over {\rm d}\theta} \Bigl[ {F_2(\theta) \over g^{\beta +2}(\theta)} 
\Bigr], \cr} \eqno\last b$$
with
\eqnam{\integralitheta}
$$I(\theta) = {g^{\beta + 2}(\theta) \over \sin^2\theta} \!
\int\limits_0^\theta \! \d\theta {\sin^2\theta \over g^{\beta+2}(\theta)} 
\bigl[ \cot\theta - {g'(\theta) \over 2g(\theta)} \bigr] h(\theta),
\eqno\new$$
and
\eqnam{\integraljtheta}
$$\eqalign{ J(\theta) =& {(\beta+\gamma -\!2)g^{\beta + 2} (\theta)
\over\sin^2\theta } \! \quad\times \cr
&\int\limits_0^\theta \! \d\theta {\sin^2\theta \over g^{\beta
+ 2}(\theta)} \bigl[ F_2(\theta)\! -\! F_1(\theta) \cot \theta \bigr].\cr} 
\eqno\new$$
Hence, the stresses can be found by evaluation of the integrals for
$I(\theta)$ and $J(\theta)$. It follows that
$$F_3(0)= F_4(0) = \fr12 h(0) - \fr12 (\beta +\gamma - 2)F_1(0).\eqno\new$$
This means that $\langle v_\theta^2 \rangle = \langle v_\phi^2
\rangle$ on the minor axis, as is required by elementary symmetry
arguments (e.g., Bacon 1985). Let us note that the case $\gamma + 
\beta =2$ is special, as then the integral $J(\theta)$ drops out of
the solution (\generalsolution). The only self-consistent model
($\gamma = 2 + \beta$) for which this happens has $\beta = 0$ and
$\gamma = 2$. This completes our derivation of the general solution of
the Jeans equations for galaxies with scale-free potentials (\genpot)
and scale-free densities of the form (\genden). The condition that the
principal components of the stress tensor are positive definite is
given in section 2.4 of ZES.

\subsection{2.3 The two-integral limit: $f=f(E, L_z^2)$} 

Let us consider briefly the special case corresponding to a
two-integral DF $f=f(E, L_z^2)$, in which the stellar velocities have
no preferred direction in the meridional plane, i.e., $\langle v_r^2
\rangle \equiv \langle v_\theta^2 \rangle$ and $\langle v_r v_\theta
\rangle \equiv 0$. It is straightforward to find the associated
stresses solution by taking
\eqnam{\twointegralchoice}
$$F_1(\theta) \equiv F_3 (\theta), \qquad\qquad 
  F_2(\theta) \equiv 0.      \eqno\new$$
Substitution of (\twointegralchoice) in equations (\generalsolution a)
and (\integraljtheta) results in a Volterra integral equation for
$F_1$ which is solved to give
\eqnam{\twointegralfone}
$$F_1 (\theta)  = {g^{\beta +2}(\theta) \over \sin^{\beta + \gamma}\theta} \!
   \int\limits_0^\theta \! \d\theta \, {\sin^{\beta + \gamma}\theta \over
 g^{\beta + 2} (\theta)} \bigl[ \cot\theta - {g'(\theta) \over 2g(\theta)}
\bigr] h(\theta).            \eqno\new$$
Substitution of this expression in equation (\generalsolution b) provides
$F_4(\theta)$ as:
\eqnam{\twointegralffour}
$$F_4(\theta) = h(\theta) + (1-\beta - \gamma) F_1(\theta).\eqno\new$$
These results can be checked using Hunter's (1977) solution of the
Jeans equations for general $f(E, L_z^2)$ models.

\begintable{1}%
 \caption{{\bf Table 1.} Coefficients for the Jeans solution (2.20), with 
           $Q=q^{-2}$.}
 \halign{#\hfil&\quad#\hfil\quad\cr
\noalign{\hrule}
\noalign{\vskip0.3truecm}
$A_{1}$         & $H_1(1+\beta-\beta q^2)+H_2(1+\beta)+\fr12(1-\beta)$\cr
\noalign{\vskip0.2truecm}
$C_{1}$         & $Q H_2(1+\beta) -\beta H_1+ \fr12 Q[2 - Q(1+\beta)]$ \cr
\noalign{\vskip0.2truecm}
$B_{2}$         & $H_1(1+\beta)$ \cr
\noalign{\vskip0.2truecm}
$A_{3}$         & $\fr12(1-\beta) - \beta q^2H_1$ \cr
\noalign{\vskip0.2truecm}
$C_{3}$ & $H_1 + \fr12 Q(2 - Q(1+\beta))$ \cr
\noalign{\vskip0.2truecm}
$A_{4}$ & $ Q(1+\beta) - \fr12(1+3\beta) - H_2(1+\beta)(1+2\beta)$ \cr
${}   $ & $\qquad  +\beta H_1 [q^2(1+2\beta) -2(1+\beta)]$\cr
\noalign{\vskip0.2truecm}
$C_{4}$         & $Q H_2(1+\beta) -\beta H_1+ \fr12 Q[2 - Q(1+\beta)]$ \cr
\noalign{\vskip0.1truecm}
\noalign{\hrule}
}
\endtable

\subsection{2.4 A two-parameter solution for the self-consistent
power-law models}

Let us now specialise to the specific case of the self-consistent
power-law models.  We take $g(\theta)$ and $h(\theta)$ as given in
(\powergtheta) and (\powerhtheta), while $\gamma = 2 + \beta$. Evans
\& de Zeeuw (1994) showed that when the stresses have the simple form
\eqnam{\simpleform}
$$\rho \langle v_j^2 \rangle = {a_j R^2 + b_j Rz + c_j z^2 \over
(R^2 + z^2/q^2)^{2+\beta}}, 
                                                                 \eqno\new$$
then the line-of-sight projected second moments -- such as the
dispersions in the radial velocities and the proper motions -- can be
explicitly evaluated as elementary functions. As a matter of choice,
we elect to study the subset of the general solution
(\generalsolution) that possesses this fetching property. This implies
choosing
\eqnam{\specialcase}
$$\eqalign{
F_1(\theta) &= I(\theta) + J(\theta) + H_1 + H_2 \, g(\theta) \sin^2\theta,\cr
F_2(\theta) &= H_2 \, g(\theta) \sin\theta \cos\theta.\cr}      \eqno\new$$
Here, $I(\theta) + J(\theta)$ is 
$$\eqalign{I(\theta)+J(\theta) =& {1\over 2}{1-\beta \over 1 + \beta}
\sin^2 \theta -{\beta \over 1 + \beta}H_1 q^2 g(\theta)\cr
&\qquad + {1\over 2}{Q [2-Q(1+\beta)]\over 1+ \beta} \cos^2 \theta,\cr
}\eqno\new$$
while $H_1$ and $H_2$ are constants. Of course, this is a very
restricted subset of the general solution (\generalsolution) --
instead of two free {\it functions} $F_1(\theta)$ and $F_2(\theta)$,
we have merely two free {\it parameters} $H_1$ and $H_2$. They have a
simple physical interpretation. $H_1 =0$ means that the velocity
ellipsoids are aligned with the cylindrical coordinate system, while
$H_2=0$ means that they are aligned with the spherical coordinate
system.

Substitution of the choice (\specialcase) in the general solution
(\generalsolution) yields
\eqnam{\fthreefourspecial}
$$\eqalign{
F_3(&\theta) = I(\theta) + J(\theta) + H_2 \, g(\theta) \cos^2\theta,  \cr
F_4(&\theta) = h(\theta) - I(\theta) - J(\theta) - 2\beta F_1(\theta) \cr
& + H_2 [Q\cos^2\theta\!-\!\sin^2\theta -2\beta (1-Q) \sin^2\theta \cos^2
      \theta ]. \cr}\eqno\new$$
This seems complicated, but when we write the Jeans solution out explicitly 
in cylindrical coordinates, it is simply:
\eqnam\wynseqa
$$\eqalign{
\rho \langle v_R^2 \rangle 
  &= {A_1 R^2 + C_1 z^2 \over (1 + \beta)(R^2 + z^2/q^2)^{2+ \beta}}, \cr
\rho \langle v_{R}v_{z}\rangle 
  &= {B_2 R z\over (1 + \beta)(R^2 + z^2/q^2)^{2+\beta}}, \cr
\rho \langle v_z^2 \rangle 
  &= {A_3 R^2 + C_3 z^2 \over (1 + \beta) (R^2 + z^2/q^2)^{2+\beta}}, \cr
\rho \langle v_\phi^2 \rangle 
  &= {A_4 R^2 + C_4 z^2 \over (1 + \beta) (R^2 + z^2/q^2)^{2+\beta}}, \cr}        \eqno\new$$
where the relationships between the coefficients $A_j, B_j, C_j$ and
our two anisotropy parameters $H_1$ and $H_2$ are given in Table 1.
Lastly, let us remark that the two-parameter Jeans solution (\wynseqa)
admits an extension to cored models that is discussed in Appendix A.

\beginfigure{1}

\centerline{\psfig{figure=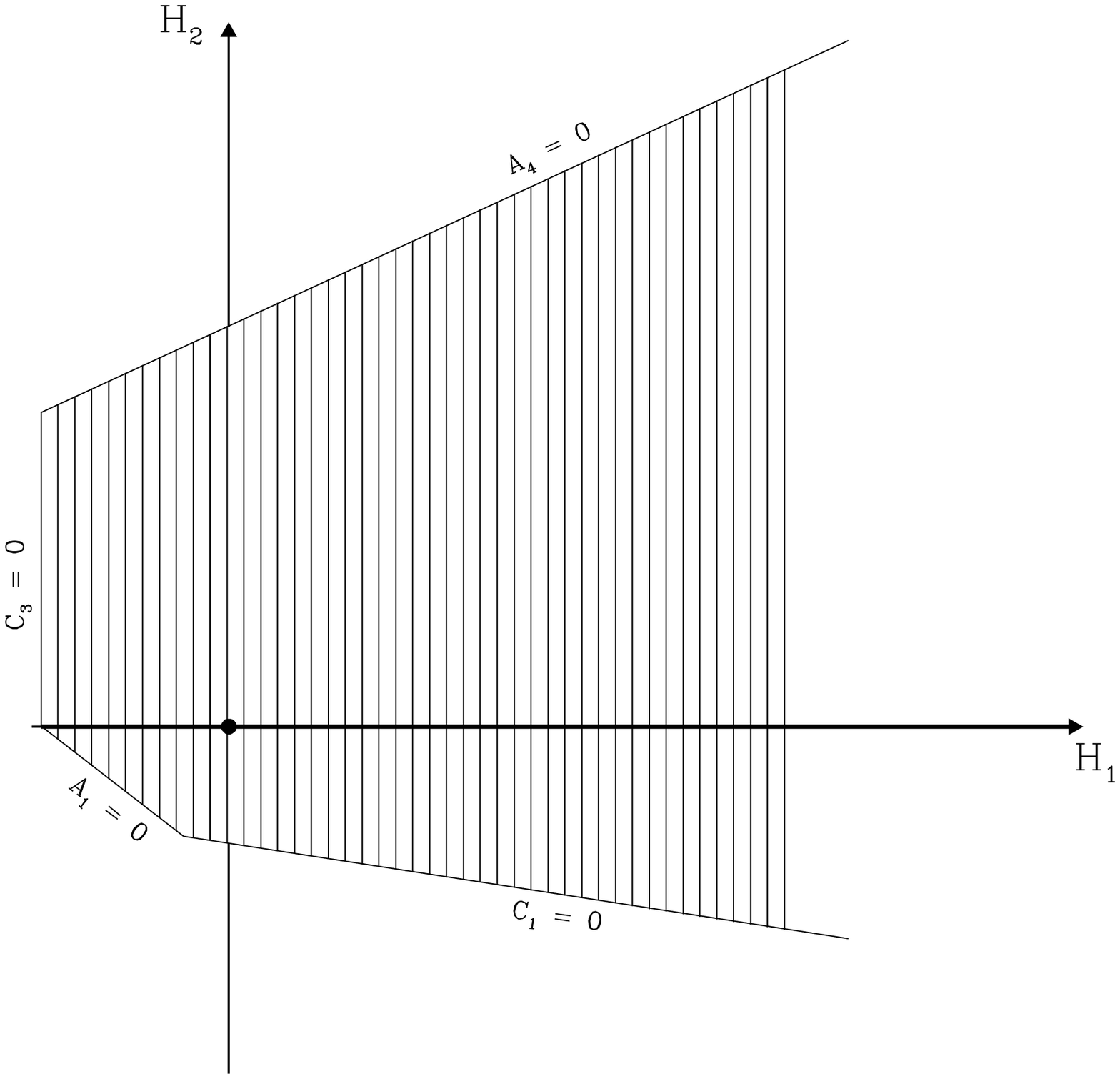,height=\hssize,width=\hssize}}

\caption{{\bf Figure 1.} The $(H_1, H_2)$-plane of two-parameter 
anisotropic solutions of the Jeans equations for the self-consistent
power-law models with $\beta = -0.25$ and $q = 0.87$. The hatched area
corresponds to the solutions of the Jeans equations with non-negative
stresses $\rho\langle v_R^2\rangle$, $\rho\langle v_z^2\rangle$, and
$\rho\langle v_\phi^2\rangle$. The requirement that the principal
components of the stress tensor $\rho\langle v_\lambda^2\rangle$ and
$\rho\langle v_\nu^2\rangle$ be positive definite leads to two further
small areas in the bottom left quadrant being deleted. The line $H_1
=0$ gives the cylindrically aligned models, while the line $H_2 =0$
gives spherically aligned models.}
\endfigure

\beginfigure{2}

\centerline{\psfig{figure=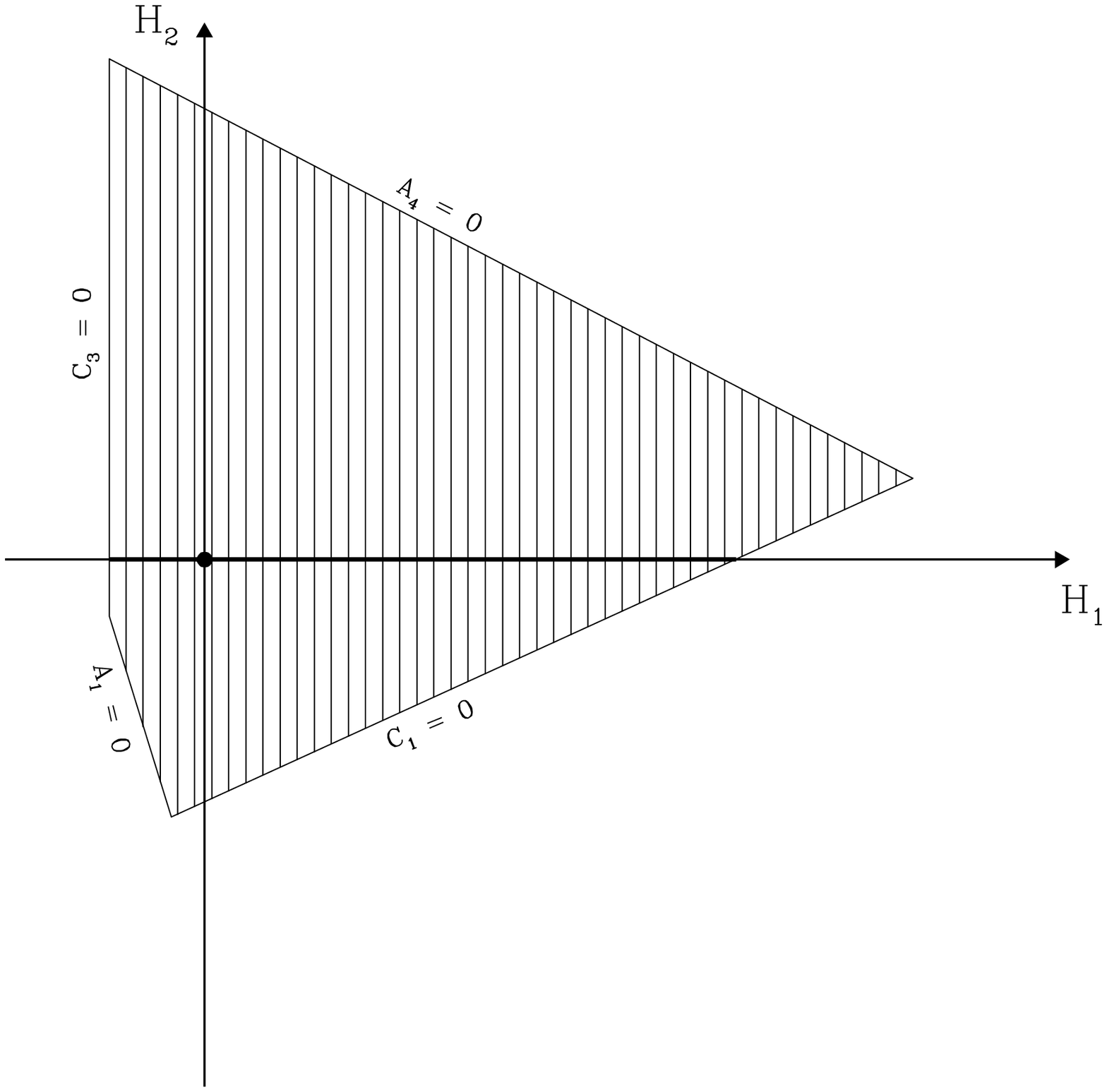,height=\hssize,width=\hssize}}

\caption{{\bf Figure 2.} As Fig. 1, but for $\beta = 0.18$ and $q= 0.9$.
The hatched area now closes, and it is not possible to find Jeans
solutions with arbitrarily large radial anisotropy. Note that the
largest value of $H_1$ consistent with positive stresses is $Q(1 -2\beta
+q^2)/(4 \beta)$. This occurs when the lines $C_1 =0$ and $A_4 =0$
intersect. Adding in the additional condition that the solution is
spherically aligned ($H_2 = 0$) gives the more stringent condition 
$\beta H_1 < Q(1 - \fr12 Q(1+\beta) )$.}
\endfigure

\subsection{2.5 Limits on the two-parameter solution} 

The principal stresses $\rho \langle v_\lambda^2 \rangle$ and $\rho
\langle v_\nu^2 \rangle$ of the two-parameter Jeans solution for
the power-law models are 
\eqnam{\principalstresses}
$$\rho \langle v_\lambda^2 \rangle = {F_-(\theta) \over r^\bandg
g^2(\theta)},
\qquad\qquad
\rho \langle v_\nu^2 \rangle = {F_+(\theta) \over r^\bandg g^2(\theta)},
                                                   \eqno\new$$
with
$$\eqalign{
F_\pm(\theta) &= I(\theta) + J(\theta) + \fr12 H_1 + \fr12 H_2 \, g(\theta) \cr
              &\pm \sqrt{H_1^2 - 2H_1 H_2 \, g(\theta) \cos 2\theta 
                              +H_2^2 g^2(\theta)}. \cr}   \eqno\new$$
The tilt angle $\Theta$ of the velocity ellipsoid in the meridional
plane is the misalignment with respect to the spherical polar coordinate
surfaces. It is:
$$\tan 2\Theta =  {2F_2 \over F_1 -F_3}= {H_2 \, g(\theta) \sin 2\theta \over
                  H_1\! -\! H_2 \, g(\theta) \cos 2\theta}.\eqno\new$$
Necessary (but {\it not} sufficient) requirements for positive
stresses are that $\rho \langle v_R^2 \rangle$, $\rho \langle v_z^2
\rangle$ and $\rho \langle v_\phi^2 \rangle$ are non-negative. This
implies the following conditions on $H_1$ and $H_2$:
\eqnam{\firstconditions}
$$\eqalign{
  -\fr12 Q[2-Q(1+\beta)] &\leq H_1, \cr
  \fr12 Q(1+\beta) -1 &\leq (1+\beta)H_2 -\beta q^2 H_1 , \cr
  -\fr12 (1-\beta) &\leq (1+\beta) H_2 \cr 
    &+ (1 + \beta - \beta q^2) H_1,\cr
    Q(1+\beta)-\fr12(1+3\beta) &\geq H_2(1+\beta)(1+2\beta)\cr
&-\beta [q^2(1+2\beta) - 2-2\beta]H_1.\cr}    \eqno\new$$
These define a hatched region in the $(H_1, H_2)$-plane illustrated in
Figs.~1 and~2 for models with rising and falling rotation curves
respectively.  Note that when $\beta <0$, the hatched region extends
to infinity along the $H_2 =0$ axis. When $\beta >0$, there exists a
limiting value of $H_1$, namely $Q(1 -2\beta +q^2)/(4 \beta)$, beyond
which the Jeans solutions become negative. The case $\beta =0$
discussed in ZES is intermediate, in the sense that the upper and
lower boundaries of the hatched region are exactly horizontal.

Necessary {\it and} sufficient conditions for positive stresses are
that the principal components $\rho \langle v_\lambda^2 \rangle$ and
$\rho \langle v_\nu^2 \rangle$ (or, equivalently, their sum and
product) are non-negative. This problem can be solved using the same
methods as in section 2.7 of ZES. The additional regions that must be
discarded from Figs.~1 and~2 are those simultaneously satisfying the
four inequalities written out in detail in Appendix B. In fact, this
results in the removal of only two small areas near the two corners of
the hatched region that lie in the third quadrant of Figs.~1 and~2, 
respectively. They lie away from the spherically aligned Jeans solutions
($H_2=0$), for which we will construct approximate DFs in Section 5.
As these additional, tiny forbidden areas are not particularly
important for the purposes of this paper, we have not complicated
Figs.~1 and~2 by marking them.

The position of the two-integral solution (\twointegralchoice) is
marked in Figs.~1 and~2 -- it lies at the origin in the $( H_1,
H_2)$-plane.  As shown in E94, it is generated by a positive definite
DF provided the flattening satisfies $q^2 \ge \fr12 (1 + \beta)$. Note
that this constraint can also be deduced from the first equation of
(\firstconditions) on putting $H_1 =0$.  It is clear from Figs.~1
and~2 that the set of Jeans solutions for the self-consistent
power-law models is large. The rest of the paper addresses the
question: which solutions are physically relevant and correspond to
positive DFs?

\eqnumber =1
\def\chaphead{\hbox{3.}}
\section{3 Spherical Potentials}

Now let us suppose the potential is exactly spherical. The beauty of
this assumption is that it enables the construction of flattened
components with simple three--integral DFs and triaxial
kinematics. This is because the total angular momentum $L$ is an exact
integral of motion in a spherical potential well, in addition to the
angular momentum component about the symmetry axis $L_z$. As first
realised by White (1985), such DFs give realistic descriptions of the
kinematics of tracer populations of stars in the outer reaches of
galaxies.

\subsection{3.1 General Jeans Solutions and Distribution Functions} 

If the gravity field is spherically symmetric, then $g(\theta)\equiv 1$
and (\genpot) reduces to the potential of the singular power-law
spheres (E94). The work in the previous section allows us to deduce
the general solution for the stresses associated with a flattened
density $\rho= r^{-\gamma} h(\theta)$. It is of the form
(\stressansatz) with $g(\theta)\equiv 1$. The functions $F_1(\theta)$
and $F_2(\theta)$ are arbitrary, while $F_3(\theta)$ and $F_4(\theta)$
follow from (\generalsolution) with $g(\theta)\equiv1$. However, in a
spherical potential, we must have $\langle v_r v_\theta \rangle
\equiv 0$ (see e.g., ZES). So, only the solutions with $F_2 (\theta) 
\equiv 0$ are physical! They have the form
\eqnam{\spherpotflatden}
$$\eqalign{
F_3(\theta) &=  I(\theta)\! +\! J(\theta), \cr
F_4(\theta) &= h(\theta) \!-\!I(\theta) \!-\! J(\theta) \!-\! 
              (\beta + \gamma \!-\!2)F_1(\theta), \cr}       \eqno\new$$
and  
$$\eqalign{
I(\theta) &= {1 \over \sin^2\theta} \int\limits_0^\theta 
              h(\theta) \sin\theta \cos\theta \, \d\theta, \cr
J(\theta) &= {(2\!-\!\beta - \gamma\!) \over \sin^2 \theta} \int_0^\theta
          F_1 (\theta) \sin\theta \cos\theta \, \d\theta. \cr}      \eqno\new$$
Here, $F_1(\theta)$ -- or the angular variation of the radial
velocity dispersion -- is arbitrary. Once it and $h(\theta)$ have been
chosen, the other second moments are fixed. When $\beta +
\gamma=3$, we have $\langle v^2 \rangle \equiv \langle v_r^2 \rangle +
\langle v_\theta^2 \rangle + \langle v_\phi^2 \rangle \equiv
r^{-\beta}$. A special case of this rule was noted earlier by Maoz \&
Bekenstein (1990), who pointed out that if the velocity dispersion is
independent of both $r$ and $\theta$, then the potential is logarithmic 
($\beta =0$) and the density profile must fall like $r^{-3}$ (i.e.,
$\gamma =3$).

\beginfigure{3}

\centerline{\psfig{figure=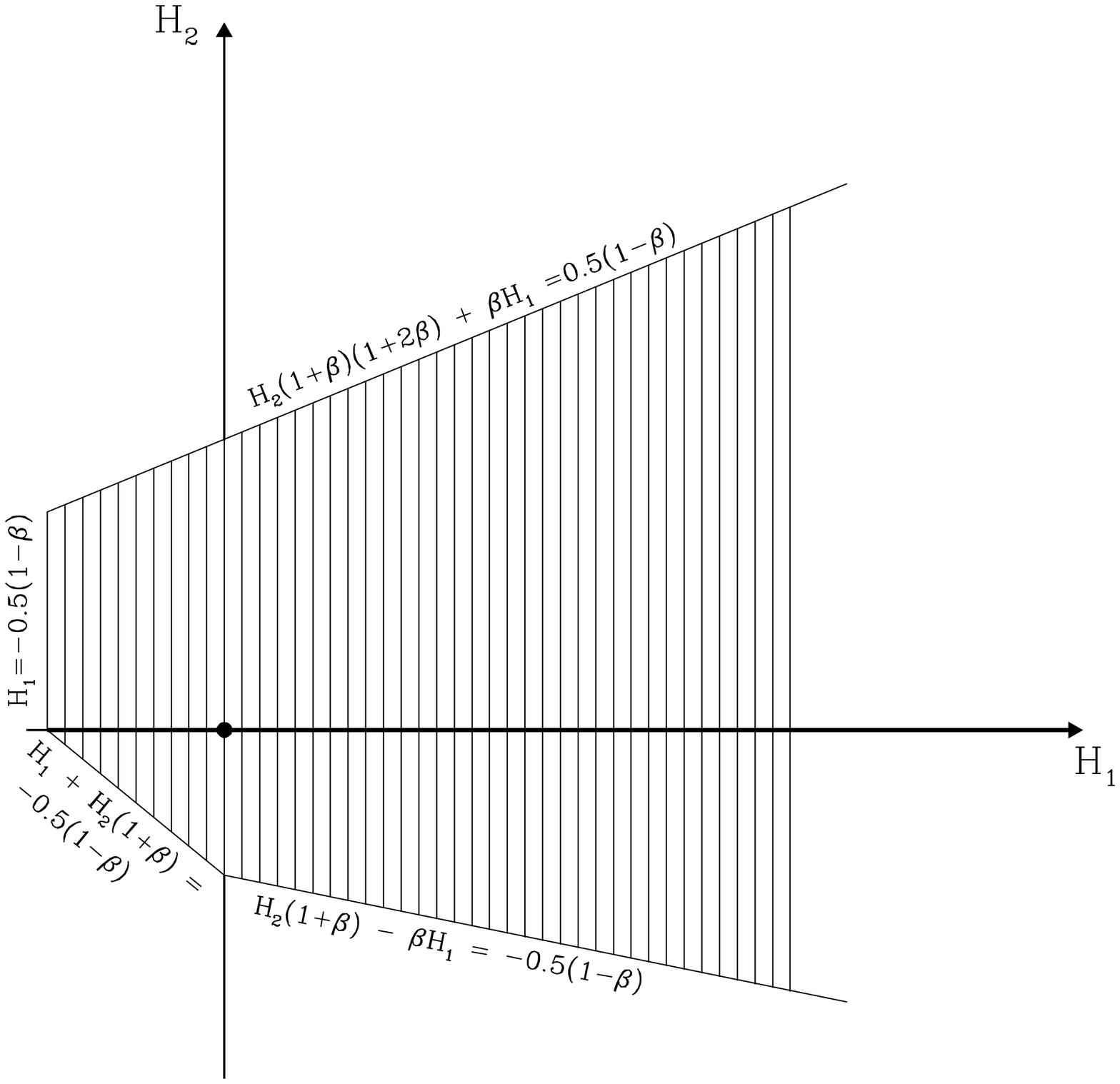,height=\hssize,width=\hssize}}

\caption{{\bf Figure 3.} The $(H_1, H_2)$-plane of two-parameter
anisotropic solutions of the Jeans equations for the singular
self-consistent power-law spheres with rising rotation laws ($\beta <
0$). The hatched region indicates the area with everywhere positive
stresses.  The filled circle corresponds to the isotropic model ($H_1
=0, H_2 = 0$). The physical solutions are those that are spherically
aligned -- these lie on the thickened, bold line $H_2 =0$, subject
only to the condition that $H_1 > - \fr12 (1-\beta)$.}
\endfigure

Now $h(\theta)$ and $F_1(\theta)$ are even functions symmetric about
$\theta = \pi/2$. They may be expanded as a series of even powers of
$\sin\theta$ (see e.g., Mathews \& Walker 1964; ZES):
\eqnam{\angvariation}
$$h(\theta)   = \sum\limits_{n=0}^\infty a_n \sin^{2n}\theta, \qquad
  F_1(\theta) = \sum\limits_{n=0}^\infty b_n \sin^{2n}\theta.     \eqno\new$$
Our aim is to find DFs generating the Jeans solution
(\spherpotflatden) corresponding to each pair $h(\theta),
F_1(\theta)$. First, we note that the stellar density law $r^{-\gamma}
\sin^{2n}\theta$ can be reproduced by non-negative DFs of the form
\eqnam\wynscomponents
$$f_{m,n} (E,L^2,L_z^2) = \eta_{m, n} L^{2m} L_z^{2n} |E|^{\alpha},
\eqno\new$$
where 
$$\alpha = {(2-\beta)(m + n)\over \beta} + {\gamma \over \beta}
- {3\over 2}.\eqno\new$$
and $m+ n > -1$, $2n > -1$ and 
$$\gamma + 2m +2n > \cases{ 0, & if $\beta <0$, \cr
          \beta (m+n +\fr12),& if $\beta >0$.\cr}
\eqno\new$$
Note that we have written $m$ and $n$ so that our notation is
consistent with White (1985) and ZES, but $m$ and $n$ are not
necessarily integers. The normalization constant $\eta_{m,n}$ is
\eqnam{\defetamn}
$$\eqalign{\eta_{m,n} =&{|\beta|^{2m/\beta + 2n/\beta + \gamma/\beta} \over
\sqrt{\pi} 2^{m+n+ 3/2}\Gamma (m+n+1) B(1/2, n+ 1/2)}\cr
& \times 
\cases{ {\displaystyle \Gamma(-\alpha)
\over \displaystyle \Gamma(-\gamma/\beta - 2m/\beta -2n/\beta )},
&if $\beta < 0$,\cr
&\null\cr &\null\cr
{\displaystyle \Gamma( \gamma/\beta  + 2m/\beta + 2n/\beta +1) \over 
\displaystyle \Gamma (\alpha+1)},& if $\beta > 0$.\cr}\cr}\eqno\new$$
Here, $B(x,y) = \Gamma(x) \Gamma(y) /\Gamma(x+y)$ is the beta function.

\beginfigure{4}

\centerline{\psfig{figure=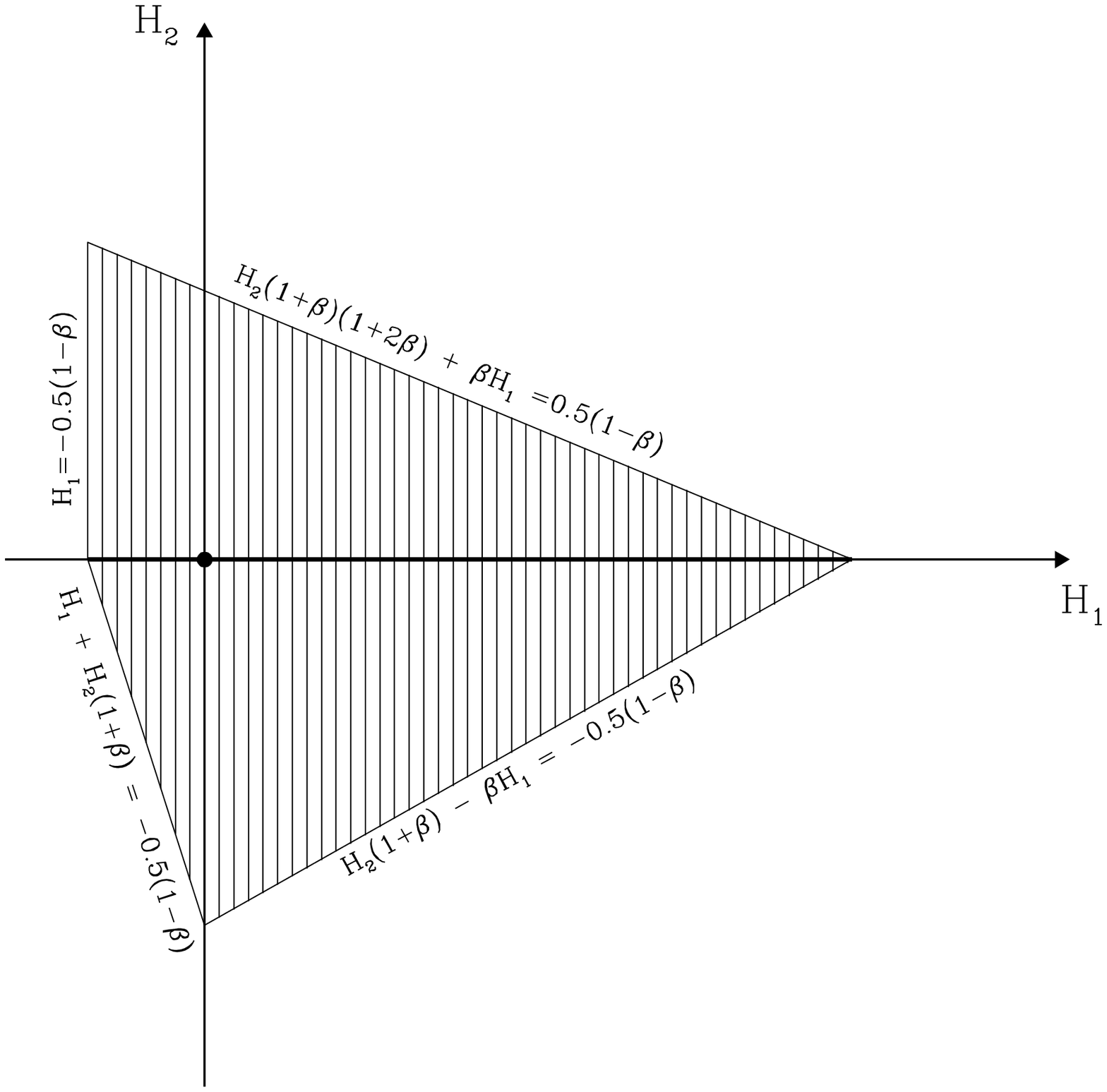,height=\hssize,width=\hssize}}

\caption{{\bf Figure 4.} As Fig.~3, but for the self-consistent 
power-law spheres with falling rotation laws ($\beta > 0$). Note that
the physical solutions (corresponding to the thickened, bold line) are
more restricted than in Fig.~3. They encompass those parts of the line
$H_2=0$, subject to $-\fr12(1-\beta) \le H_1 < \fr12
(1-\beta)/\beta$. }
\endfigure

\noindent
By integrating over all velocity space, the corresponding stresses are
\eqnam{\wyns}
$$\eqalign{
\rho \langle v_r^2 \rangle &= {1 \over 2m + 2n + \bandg }
                       {\sin^{2n}\theta \over r^\bandg}, \cr 
\rho \langle v_\theta^2 \rangle &= { m + n + 1 \over n+1} 
                        \rho\langle v_r^2 \rangle, \cr 
\rho\langle v_\phi^2 \rangle &= (2n+1) \rho\langle v_\theta^2 \rangle.
\cr} \eqno\new$$
The angular dependence of the density is independent of $m$ for fixed
$\gamma$ and $n$. The same is true of the stresses.  However, the
anisotropy ratios $\langle v_\theta^2 \rangle/ \langle v_r^2 \rangle$
and $\langle v_\phi^2 \rangle/\langle v_r^2 \rangle$ do vary with
$m$. Whenever $\gamma + \beta = 3$, then $\langle v_r^2 \rangle +
\langle v_\theta^2 \rangle + \langle v_\phi^2 \rangle = v^2_{\rm
circ}$. In the limit $\beta \rightarrow 0$, the components reduce to
those studied in White (1985), Gerhard (1991) and ZES. The components
for $\beta >0$ have previously been discussed by Kulessa \&
Lynden-Bell (1992) and de Bruijne, van der Marel \& de Zeeuw
(1996). Using the method of linear superposition of the components
(\wynscomponents), the general solution of the Boltzmann equation
becomes
\eqnam{\generaldfsphere}
$$f(E, L^2 , L_z^2) = \sum_{m,n} A_{m,n}f_{m,n}.            \eqno\new$$
Here, the $A_{m,n}$ are constants specifying the fraction
contributed by each component. To build DFs that recover the angular
variation $h(\theta)$ of the density and $F_1(\theta)$ of the radial 
velocity dispersion of our Jeans solution, we must insist that:
$$\sum\limits_{m=0}^\infty A_{m,n} = a_n, \qquad\qquad
 \sum\limits_{m=0}^\infty {A_{m,n} \over 2m\!+\!2n\!+\bandg} =
b_n,\eqno\new$$
for $n=0, 1, 2, \ldots$. There are many more unknowns $A_{m,n}$ than
constraints $a_n, b_n$. So, there are many ways to define
three-integral DFs that reproduce the required density and velocity 
dispersions. For specific choices of $h(\theta)$, examples of
such DFs are given in Kulessa \& Lynden-Bell (1992) and de Bruijne,
van der Marel \& de Zeeuw (1996). 

As an aside, we remark that it is possible to generalise (\wynscomponents) 
to produce triaxial stellar distributions with triaxial kinematics in 
spherical potentials. In view of its ready application to the kinematics
of the stellar halo of the Milky Way Galaxy, this development is 
discussed in more detail in Appendix C.

\begintable*{2}%
 \caption{{\bf Table 2.} Accuracy of the partial integral $I_3$ for
an E3 power-law model with a falling rotation curve ($\beta =0.18, q=
0.9$). Meridional cross sections of the orbits are displayed in
Fig.~5.}
 \halign{#\hfil&#\hfil&\hfil#\hfil&\hfil#\hfil&
   $\quad$#\hfil&\quad\hfil#\hfil&\hfil#\hfil&\hfil#\hfil&$\quad$\hfil#
&\hfil#\hfil&\hfil#\hfil\cr
\noalign{\hrule}
\noalign{\vskip0.1truecm}
 {}  & $q$  & $R_0$  & $z_0$  & $L_z$
     & $I_{3, \rm max}$ & $I_{3, \rm min}$ & $\delta I_3$
     & $L^2_{\rm max}$  & $L^2_{\rm min}$  & $\delta L^2$  \cr
\noalign{\vskip0.3truecm}
\noalign{Thin tubes in E3 galaxies}
\noalign{\vskip0.1truecm}
$A$  & 0.9  & .5232  & .2384  & .541  & .3678  & .3677 & .0001
                                      & .3677  & .3534 & .0143 \cr
$B$  & 0.9  & .3794  & .4268  & .378  & .3671  & .3660 & .0011
                                      & .3660  & .3236 & .0424 \cr
$C$  & 0.9  & .1889  & .5367  & .182  & .3657  & .3630 & .0027
                                      & .3630  & .3004 & .0626 \cr
$D$  & 0.9  & .0472  & .5667  & .045  & .3651  & .3617 & .0034
                                      & .3617  & .2935 & .0682 \cr
\noalign{\vskip0.2truecm}
\noalign{Fat tubes in E3 galaxies}
\noalign{\vskip0.1truecm}
$E$  & 0.9  & .4800  & .4853  & .378  & .3499  & .3208 & .0291
                                      & .3484  & .2889 & .0595 \cr
$F$  & 0.9  & .6600  & .4529  & .378  & .2583  & .2208 & .0375
                                      & .2531  & .2094 & .0437 \cr
$G$  & 0.9  & .8400  & .2425  & .378  & .1666  & .1555 & .0111
                                      & .1650  & .1539 & .0111 \cr
\noalign{\vskip0.1truecm}
\noalign{\hrule}
}

\endtable

\subsection{3.2 A one-parameter solution for self-consistent spherical
models}

Let us now consider the limiting behavior of the two-parameter Jeans
solutions for the self-consistent power-law spheres. Not just
the potential, but also the density, is now spherical.  Using
$h(\theta) \equiv 1-\beta$, $g(\theta) \equiv 1$ and $\gamma=2+\beta$, 
we find from equations (\specialcase) that
$$\eqalign{
F_1(\theta) &= {1 - \beta  + 2H_1\over 2(1 + \beta )} 
+ H_2\sin^2\theta, \cr
F_2(\theta) &= H_2\sin\theta\cos\theta, \cr
F_3(\theta) &= {1 - \beta -2\beta H_1 \over 2(1 + \beta )} +  
H_2\cos^2\theta, \cr
F_4(\theta) &= {1 - \beta -2\beta H_1 \over 2(1 + \beta)} +  
H_2[\cos^2\theta -\sin^2\theta(1+2\beta)]. \cr}      \eqno\new$$
These anisotropic stresses for the self-consistent, singular power-law
spheres are positive in the region in the $(H_1, H_2)$-plane defined
by (cf.\ eq.\ [\firstconditions])
$$\eqalign{
 -\fr12(1-\beta)  &\leq H_1, \cr
 -\fr12(1-\beta) &\leq H_2(1+\beta)-\beta H_1, \cr
 -\fr12(1-\beta) &\leq H_2(1+\beta) + H_1, \cr
 \fr12(1-\beta) &\geq H_2(1+\beta)(1+2\beta) +\beta H_1,   \cr}                                     \eqno\new$$
which is illustrated in Figs.~3 and~4 for power-law spheres with rising
($\beta <0$) and falling ($\beta >0$) rotation laws. As we have seen, 
all physical solutions must have $F_2 \equiv 0$ because of the properties 
of individual orbits. On requiring that $H_2=0$, we are left with a
one-parameter solution
$$\eqalign{
F_1(\theta)&=  {1 - \beta +  2H_1\over 2(1 + \beta )}, \cr
F_3(\theta)&= F_4(\theta)= {1 - \beta -2\beta H_1\over 2(1 + \beta)},\cr} 
\eqno\new$$ 
which is indicated as the thick solid line in Figs.~3 and~4. It follows 
from the derivation in Section 3.1 that each of these Jeans solutions
corresponds to a DF of the general form (\generaldfsphere) with $n=0$,
so that  
\eqnam{\felsquared}
$$f(E, L^2) = \sum_{m=0}^\infty A_{m,0} \eta_{m,0} 
               L^{2m} |E|^{2(m+1)/\beta -m -1/2},                \eqno\new$$ 
with $\eta_{m,0}$ given in equation (\defetamn). The coefficients are
subject to the constraints 
$$\sum\limits_{m=0}^\infty A_{m,0} = 1 - \beta, \qquad
\sum\limits_{m=0}^\infty {A_{m,0} \over m\!+\!\beta\!+\!1} = 
{1-\beta + 2H_1 \over 1 + \beta}. 
                                                           \eqno\new$$
Again, many DFs are possible. The simplest is obtained by taking only
one component in the series (\felsquared):
\eqnam{\simplechoice}
$$f(E, L^2) = \eta_{m,0} L^{2m} |E|^{2(m+1)/\beta -m -1/2}, \eqno\new$$
with 
$$m = {-2H_1 (\beta + 1)\over 1 + 2H_1 - \beta}.    \eqno\new$$
Of course, $H_1$ is a measure of the anisotropy of the model. $H_1
=-\fr12 (1-\beta)$ corresponds to the circular orbit model (no radial
velocity dispersion) and $H_1=0$ to the isotropic model. When $\beta \le
0$, then $H_1 \to \infty$ and the model approaches the radial orbit
model. When $\beta > 0$, then the most radially distended model possible
(see Fig.~4) has $H_1 = (1-\beta)/(2\beta)$. This corresponds to the 
critical value of $m = -1$, for which (\wynscomponents) ceases to exist. 
The DF (\simplechoice) is non-negative and so physical for each of these 
permitted values of $H_1$.

\begintable*{3}%
 \caption{{\bf Table 3.} Accuracy of the partial integral $I_3$ for
          an E3 power-law model with a rising rotation curve ($\beta =-0.25,
          q= 0.87$). }
 \halign{#\hfil&#\hfil&\hfil#\hfil&\hfil#\hfil&
   $\quad$#\hfil&\quad\hfil#\hfil&\hfil#\hfil&\hfil#\hfil&$\quad$\hfil#
&\hfil#\hfil&\hfil#\hfil\cr
\noalign{\hrule}
\noalign{\vskip0.1truecm}
 {}  & $q$  & $R_0$  & $z_0$  & $L_z$
     & $I_{3, \rm max}$ & $I_{3, \rm min}$ & $\delta I_3$
     & $L^2_{\rm max}$  & $L^2_{\rm min}$  & $\delta L^2$  \cr
\noalign{\vskip0.3truecm}
\noalign{Thin tubes in E3 galaxies}
\noalign{\vskip0.1truecm}
$A$  & 0.87  & .5852  & .2573  & .541  & .3677  & .3674 & .0003
                                      & .3674  & .3493 & .0181 \cr
$B$  & 0.87  & .4334  & .4677  & .378  & .3653  & .3628 & .0025
                                      & .3627  & .3094 & .0533 \cr
$C$  & 0.87  & .2220  & .6003  & .182  & .3600  & .3527 & .0073
                                      & .3527  & .2753 & .0774 \cr
$D$  & 0.87  & .0562  & .6395  & .045  & .3566  & .3477 & .0089
                                      & .3477  & .2640 & .0837 \cr
\noalign{\vskip0.2truecm}
\noalign{Fat tubes in E3 galaxies}
\noalign{\vskip0.1truecm}
$E$  & 0.87  & .4800  & .4961  & .378  & .3678  & .3438 & .0240
                                      & .3676  & .2955 & .0721 \cr
$F$  & 0.87  & .6600  & .4927  & .378  & .3037  & .2444 & .0593
                                      & .3010  & .2225 & .0785 \cr
$G$  & 0.87  & .8400  & .3423  & .378  & .2079  & .1742 & .0337
                                      & .2027  & .1666 & .0361 \cr
\noalign{\vskip0.1truecm}
\noalign{\hrule}
}

\endtable

\eqnumber =1
\def\chaphead{\hbox{4.}}
\section{4 The Boltzmann Approach} 

The Boltzmann approach emphasises the primacy of distribution functions
(DFs). In this section, we investigate which of the two-parameter
Jeans solutions can correspond to three-integral DFs for the flattened
power-law potentials. To do this, we must first extend the theory of
partial integrals introduced in ZES.

\subsection{4.1 Classical and partial integrals}

The collisionless Boltzmann equation for the DF $f$ can be cast into 
the form
\eqnam{\cbeab}
$$0 = v_r A + v_\theta B,                                      \eqno\new$$
with
\eqnam{\adefinition} 
$$A = \fr12 r {\partial f \over \partial r} + 
     (v_\theta^2\!+\!v_\phi^2\!-\!r {\partial \Phi \over \partial r}) 
      {\partial f \over \partial v_r^2} 
   -v_\theta^2 {\partial f \over \partial v_\theta^2} 
          -v_\phi^2 {\partial f \over \partial v_\phi^2},      \eqno\new$$
and
\eqnam{\bdefinition}
$$B = \fr12 {\partial f \over \partial \theta} 
     - {\partial \Phi \over \partial \theta} 
       {\partial f \over \partial v_\theta^2} 
  +v_\phi^2 \cot\theta 
   \Big( {\partial f \over \partial v_\theta^2} 
        -{\partial f \over \partial v_\phi^2} \Big).           \eqno\new$$
Of course, any axisymmetric potential has two classical integrals, the
energy $E=\Phi +\fr12 (v_r^2 + v_\theta^2 + v_\phi^2)$ and the
$z$-component of the angular momentum $L_z = r v_\phi
\sin\theta$. With the exception of special potentials possessing
additional symmetries in phase space, axisymmetric potentials do not
have exact, globally defined third integrals (e.g., Lynden-Bell 1962;
de Zeeuw 1985; Evans 1990).  This is awkward for stellar dynamics, as
the observational data require three-integral DFs. One way round this
problem is suggested in ZES. We look for partial integrals which have
good accuracy for some orbital families, rather than the global
integrals which we know in general do not exist.  ZES constructed such
a partial integral for the thin and near-thin tubes in a scale-free
model with a flat rotation curve (Binney's model). Here, our aim is to
understand how this partial integral extends to the entire scale-free
family.

In the spherical limit, the potential has an exact third integral, 
namely the total angular momentum $L^2 = r^2 (v_\theta^2 + v_\phi^2)$. 
ZES introduced a modification 
\eqnam{\defithree}
$$I_3= L^2 i_3 (\theta) = r^2 (v_\theta^2 + v_\phi^2) i_3 (\theta),
\eqno\new$$
and showed that a suitable choice for Binney's model was $i_3 (\theta)= 
\sin^2 \theta + Q\cos^2 \theta$. Inserting our ansatz $I_3$ into the 
Boltzmann equation (\cbeab), we find that
$$\eqalign{A=& 0,\cr
B =& \fr12 r^2 i_3'(\theta)\Bigl[v_\theta^2 +v_\phi^2 - {1\over
r^\beta} {i_3(\theta) g'(\theta)\over i_3'(\theta)g^{1 + \beta/2}(\theta)}
\Bigr].\cr}\eqno\new$$
The thin or near-thin tubes have the property that 
\eqnam{\intcond}
$$v_\theta^2 + v_\phi^2 \approx -r{\partial \Phi \over \partial r} 
= {1\over r^{\beta}g^{\beta/2}(\theta)},\eqno\new$$
is fulfilled. This essentially states that anywhere on a thin tube,
$v_r$ is much smaller than the other two components. This itself implies
that, in the meridional plane, the thin tubes lie very nearly on circles. 
This is certainly a good approximation in moderately flattened models
(like E3). The partial integral therefore satisfies the collisionless 
Boltzmann equation if
$${1\over i_3}{ {\rm d} i_3\over {\rm d} \theta} = {1\over g}{ {\rm d} g \over
{\rm d} \theta}.\eqno\new$$
Straightforward integration gives
\eqnam{\ithree}
$$i_3 (\theta) = g(\theta),\eqno\new$$
and so the third integral is just
\eqnam{\defpartint}
$$I_3 = r^2 (v_\theta^2 + v_\phi^2) g(\theta).\eqno\new$$
This is exactly the same as deduced by ZES for the specific case of
Binney's model. We emphasise, though, that the only assumption involved 
in deriving (\ithree ) is that of scale-freeness of the potential. The 
claim is that this is a good partial integral for thin and near-thin
tubes in {\it any} galaxy with a scale-free potential. 

\beginfigure{5}

\centerline{\psfig{figure=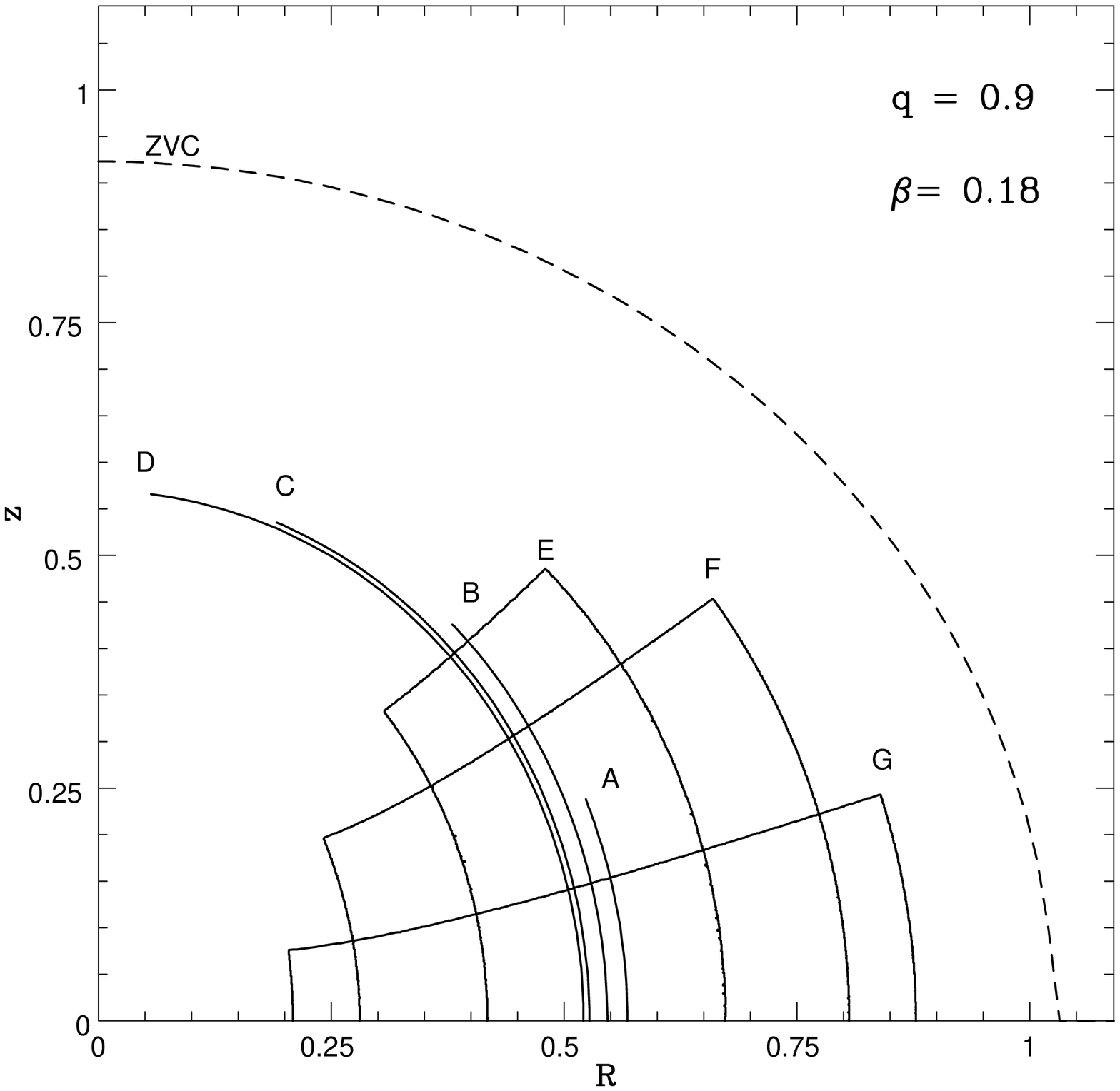,height=\hssize,width=\hssize}}

\caption{{\bf Figure 5.} Cross sections of the orbits of Table~2 with a
meridional plane $(R, z)$ for the scale-free power-law potential 
$q=0.9$ at energy $E=-5.5811$. The zero-velocity curve
(for $L_z=0$) is indicated by the ZVC. Orbits A - D are thin
tubes. Orbits E - G are fat (or thick) tubes that fill the indicated
areas. (Cross--sections of the orbits in Table~3 look very similar).} 
\endfigure

\beginfigure{6}

\centerline{\psfig{figure=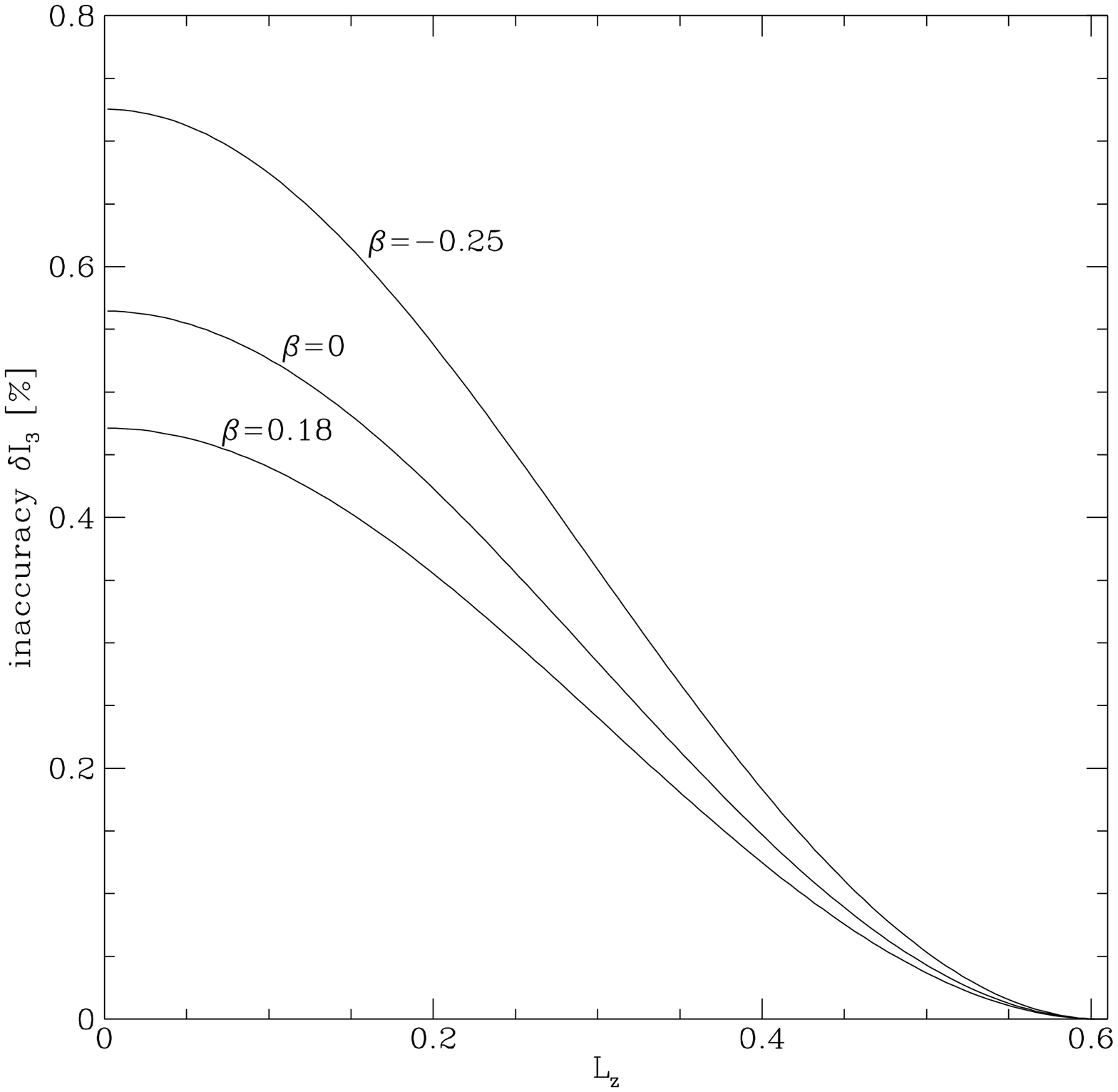,height=\hssize,width=\hssize}}

\caption{{\bf Figure 6.} The fluctuation in the partial integral
for the infinitesimally thin tubes (expressed as a percentage) is plotted 
against the angular momentum for three values of $\beta$. Even for the thin
tubes, the partial integral is not exact, although it is very good.
The diagram clearly illustrates that the partial integral worsens
in accuracy as $\beta$ diminishes. (All the models have $q =0.9$).
}
\endfigure

\subsection{4.2 Numerical integrations}

The accuracy of the partial integral is now investigated by adaptive
Runge-Kutta techniques for the specific cases of the self-consistent
scale-free power-law galaxies (i.e., $g(\theta)$ is given by
(\powergtheta)).  The results of such computations are listed in
Tables~2 and~3, while Fig.~5 shows cross-sections of the orbits in the
meridional plane. To ease comparison with the results of ZES, the
energy surface is always normalised to
\eqnam{\eref}
$$E = \fr12 {\beta -2 \over \beta} \exp \Bigl[ {\beta \over 2-\beta}
\Bigr],\qquad\qquad \beta \neq 0.\eqno\new$$
This ensures that the limiting angular momentum of the circular
orbit is $e^{-1/2}$. For the purposes of illustration, we concentrate
on one model with falling rotation curve ($\beta = 0.18, q= 0.9, E =
-5.5811$) and one model with rising rotation curve ($\beta = -0.25, q
= 0.87, E = 4.0268$). The rationale behind the choices of the
equipotential axis ratio $q$ is that the galaxies have the same
asymptotic ellipticity of E3 (using formula (2.9) of E94).  The values
of $\beta$ are suggested by taking the means of the asymptotic
logarithmic gradients of the rotation curves in the samples of
Casertano \& van Gorkom (1991). In Tables~2 and~3, the orbits are
identified by their starting values $R_0$ and $z_0$, with $v_r$ and
$v_\theta$ zero there. The maximum and minimum values of $I_3$ and
$L^2$ are recorded, together with their fluctuation. Figure~5 shows
cross-sections of the orbits for $\beta = 0.18$ (the diagram for
$\beta = -0.25$ is very similar). The thin tubes are labelled A, B, C,
and D. For all these thin tubes, the partial integral is
well-conserved. The lower section of Tables~2 and~3 shows the results
for thicker tube orbits. Now the variation in the partial integral is
larger -- but it is still more accurately conserved than the total
angular momentum.  The partial integral is not exact, even for the
thin tubes. At any angular momentum $L_z$, there is a unique 
infinitesimally thin tube with energy (\eref). Fig.~6 shows the 
fluctuation in $I_3$ (expressed as a percentage) along the thin tubes
for the three values $\beta = 0.18, 0.0$ and $-0.25$. This illustrates
graphically that the fluctuations in the partial integral are smaller 
for the models with falling rotation curves. Note that, even for
the thin tubes, the partial integral is not exact.

\beginfigure{7}

\centerline{\psfig{figure=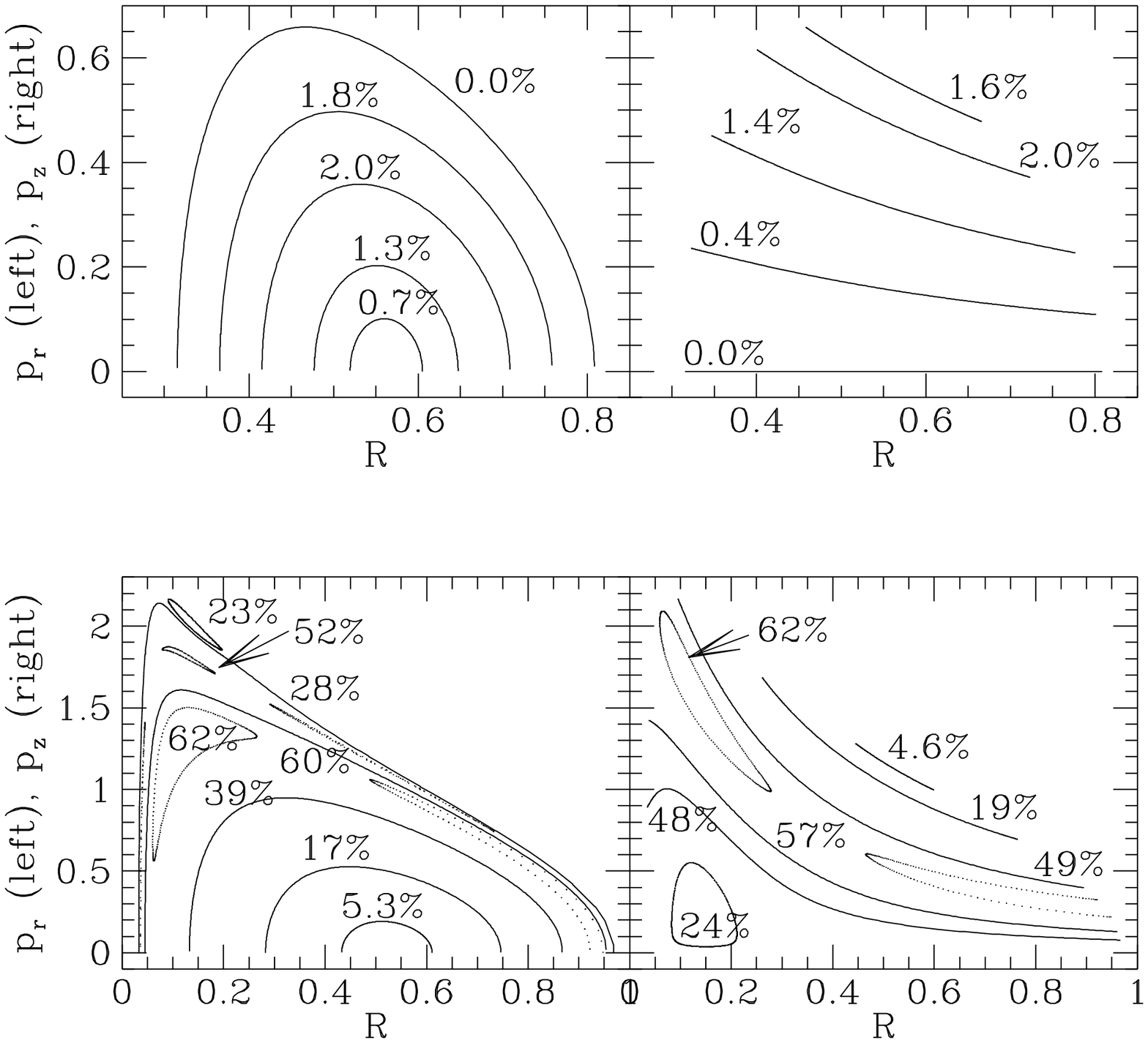,height=\hssize,width=\hssize}}

\caption{{\bf Figure 7.} Poincar\'e surfaces of section for the
model with $\beta = 0.18$ and $q = 0.9$. The energy hypersurface is
normalised to $E = -5.5811$. The upper and lower panels show the the
($R,p_R$) and ($R,p_z$) cross-sections of the energy hypersurface at
the equatorial plane for $L_z =0.5$ and at $L_z =0.1$
respectively. Each orbit is marked with the fluctuation in the partial
integral represented as a percentage.  At high $L_z$, the orbits are
mainly thin and near-thin tubes. As is evident from the upper two
panels, the partial integral is well-conserved by all the orbits. At
low $L_z$, minor orbital families trapped around the $1:1$ and $3:4$
resonances (the \lq reflected banana' and \lq reflected fish'
families) make their appearance. The lower two panels show the partial
integral is poorly conserved for these families. Near the thin-tube
orbit, though, the conservation is good.}
\endfigure

Can the partial integral accurately distinguish between different
orbital families or -- equivalently -- different tori in phase space?
One way to establish this is to investigate the phase space structure
using Poincar\'e surfaces of section (see e.g., Gutzwiller 1990).
Fig.~7 shows surfaces of section for the model with a falling rotation
curve ($\beta = 0.18, q= 0.9, E = -5.5811$). The upper and lower
panels show the ($R,p_R$) and ($R,p_z$) cross-sections at the equator
for large and small $L_z$ respectively. In each case, the orbits are
marked with the percentage fluctuation of the partial integral.  At
high $L_z$, there is the reassuring picture of good conservation of
the partial integral all across the surface of section. The orbits are
all tubes. At low $L_z$, the resonances associated with the \lq
reflected banana' and \lq reflected fish' families (see e.g. Lees \&
Schwarzschild 1992; E94) make their appearance.  The partial integral
is poorly conserved on the islands in the lower panels of Fig.~7. Only
near the thin-tube orbits and near the orbits confined to the
equatorial plane is the preservation of the partial integral at all
good in the lower-panels. It is useful to contrast this behaviour with
that of the classical integral $L^2$, which does not distinguish
between any of the orbits on the lower panel of Fig.~7. We conclude
then that $I_3$ is a useful discriminant between orbits of the same 
$E$ and $L_z$.

Although we have only presented numerical evidence for the power-law
models, which have spheroidal equipotentials, nonetheless the derivation 
given in Section 4.1 suggests that our partial integral is 
good for thin and near-thin tubes in all axisymmetric scale-free 
potentials. It would be interesting to substantiate this claim for
other scale-free galaxy models -- such as those studied by
Toomre (1982) for instance.

\eqnumber =1
\def\chaphead{\hbox{5.}}
\section{5 Three-integral distribution functions}

We now exploit the partial integral for the scale-free power-law
models to build simple three-integral DFs by superposing basic component
solutions of the collisionless Boltzmann equation. First, we develop
some preliminary results concerning our components in Section 5.1. 
Then, in the next section, we pass to the actual construction of the
approximate analytical three-integral DFs by superposition of the
components.

\subsection{5.1 Components}
 
The two exact integrals $I_1 =|E|$, $I_2 = L_z^2$ and the partial
integral $I_3$ can be used to construct component DFs $f(I_1, I_2,
I_3)$. Let us note that all such DFs correspond to spherically 
aligned Jeans solutions, i.e.,
$$\langle v_r v_\theta \rangle = \langle v_\theta v_\phi \rangle =
\langle v_r v_\phi \rangle \equiv 0.\eqno\new$$
This general result follows directly from the forms of the integrals.
In particular, it depends on the absence of velocity cross-terms in
the partial integral. Is it possible to improve the accuracy of the
partial integral by incorporating velocity cross-terms, which generate
misalignment of the velocity ellipsoid with the spherical polar
coordinates? We have explored this question with numerical orbit
integration. It appears that some slight improvement of the partial
integral is possible, but only at the cost of increasing mathematical
complexity of the partial integral. We believe that the coefficient 
of any cross-term must be small, and so any misalignment of the
stresses from spherical polars is also limited.

To build three-integral DFs, we shall find it useful to study the
properties of two sets of components. The first set is just the
product of powers of the integrals (cf., Fricke 1951; White 1985; ZES)
\eqnam{\basicsolutionsone}
$$f_{k,n,m} = I_1^k I_2^n I_3^m.                      \eqno\new$$
The density generated by (\basicsolutionsone) is
\eqnam{\basicdensityone}
$$\rho_{k,n,m} \propto {r^{ (2- \beta)(m+n) -\beta(k + 3/2) }\sin^{2n} 
\theta \, [g (\theta) ]^m \over [g(\theta)]^{\beta (k+m+n)/2 + 3\beta/4} }.
\eqno\new$$
To reproduce the $r$-dependence of the density, we must have
\eqnam{\krestrict}
$$k = {(2 - \beta ) (m+n) \over \beta} + {2\over \beta} -
{1\over 2},\eqno\new$$
so that (\basicdensityone) simplifies to
\eqnam{\basicsimpledensityone}
$$\rho_{n,m} = {S_{n,m}\over r^{2+\beta}}{\sin^{2n} 
               \theta \over [ g(\theta)]^{1+ n + \beta/2}},\eqno\new$$
with 
\eqnam{\defSmn}
$$\eqalign{&S_{m,n} ={\sqrt{\pi} 2^{m+n+ 3/2}\Gamma (m+n+1)B(1/2, n+ 1/2)
\over |\beta|^{1 + 2(m + n + 1)/\beta}}\cr
&\times
\cases{ {\displaystyle \Gamma(-1 -2(m+n+1)/\beta)
\over \displaystyle \Gamma(1/2 -2/\beta - (2-\beta) (m+n)/\beta )},
&if $\beta < 0$,\cr
&\null\cr &\null\cr
{\displaystyle \Gamma( 1/2 + 2/\beta + (2-\beta)(m+n)/\beta ) \over
\displaystyle \Gamma (2 + 2(m+n+1)/\beta)},& if $\beta > 0$.\cr}
\cr}\eqno\new$$
The second set of components is 
\eqnam{\basicsolutionstwo}
$$g_{k,n,m} = I_1^k I_2^n (I_3 + C I_2)^m.                     \eqno\new$$
Here, $C$ is a constant, which we assume to be small and check a
posteriori. Again, to reproduce the $r$-dependence of the density,
$k$ is restricted to (\krestrict). The density corresponding to
(\basicsolutionstwo) is
\eqnam{\basicsimpledensitytwo}
$$\rho_{n,m} = {S_{n,m}\over r^{2+\beta}}{\sin^{2n}
               \theta \over [ g(\theta)]^{1+ n + \beta/2}}\Bigl[
1 + C{m(2n+1)\over 2(n+1)}{\sin^2 (\theta)\over g(\theta)} \Bigr].
\eqno\new$$
This follows on taking the Taylor expansion of (\basicsolutionstwo)
and then using (\basicsimpledensityone) twice. In the spherical
limit, we shall show that $C =0$. Therefore, both sets of components
(\basicsolutionsone) and (\basicsolutionstwo) are axisymmetric
generalisations of the components (\wynscomponents) in the spherical
limit.

\subsection{5.2 Three-Integral DFs}

DFs that can reproduce the power-law density are given by 
restricting attention to just the $n=0$ and $n=1$ components in
(\basicsimpledensityone) and (\basicsimpledensitytwo). We choose
to investigate DFs of the form:
\eqnam{\approxdf}
$$\eqalign{f =& \sum_m A_{0,m} g_{k,0,m} + A_{1,m} f_{k,1,m},\cr
 =& \sum_m A_{0,m} I_1^{2/\beta+(2-\beta)m/\beta - 1/2}
(I_3 + CI_2)^{m}\cr
+& \sum_m A_{1,m} I_1^{2/\beta + (2-\beta)(m+1)/\beta - 1/2} I_2 I_3^m,\cr} 
\eqno\new$$
where the $A_{0,m}$ and $A_{1,m}$ are unknown amplitudes. This 
generates the density
\eqnam{\dfdensity}
$$\eqalign{\rho =& {1\over r^{2 + \beta}g^{2 + \beta/2}(\theta)}
     \Bigl[ g(\theta) \sum\limits_m A_{0,m}S_{0,m} \cr 
     +&  \sin^2 \theta \Bigl(\sum\limits_m  \fr12 m C A_{0,m}S_{0,m} 
     + A_{1,m}S_{1,m}\Bigr) \Bigr].\cr}\eqno\new$$
If the DF only depends on globally defined classical integrals, then
it obviously satisfies the Jeans equations. This is not guaranteed for
DFs depending on any partial or approximate integrals. So, we must also
compare the kinematics of the DFs (\approxdf) with the solutions of the
Jeans equation derived earlier in Section 2. By straightforward
integration over velocity space, we discover that the DFs (\approxdf)
generate the second moments:
\eqnam{\dfkin}
$$\eqalign{
\rho \langle v_r^2 \rangle &\!=\!
 {1 \over r^{2+ 2\beta}g^{2 + \beta}(\theta)} \Bigl[ 
     \fr12 g(\theta) \sum\limits_m
      {A_{0,m}S_{0,m} \over (1+m + \beta)} \cr
  &+\fr12 \sin^2\theta \sum\limits_m
      {A_{1,m}S_{1,m} \over (2+m + \beta)}  \Bigr],\cr
\rho \langle v_\theta^2 \rangle &\!=\!
 {1 \over r^{2+ 2\beta}g^{2 + \beta}(\theta)} \Bigl[ 
    \fr12  g(\theta) \sum\limits_m
      {(1+m)A_{0,m}S_{0,m} \over (1+m + \beta)} \cr
  &- \fr18 C\sin^2\theta 
   \sum\limits_m {m(1 +m) A_{0,m}S_{0,m}\over (1+m + \beta)}\cr 
  &+\fr14 \sin^2\theta 
        \sum\limits_m
      {(2 + m)A_{1,m}S_{1,m} \over (2+m + \beta)}  \Bigr],\cr
\rho \langle v_\phi^2 \rangle &\!=\!
 {1 \over r^{2+ 2\beta} g^{2 + \beta}} \Bigl[ 
      \fr12 g(\theta) \sum\limits_m
      {(1+m)A_{0,m}S_{0,m} \over (1+m + \beta)} \cr
  &+\fr14 C\sin^2 \theta 
      \sum\limits_m {m(1 +m) A_{0,m} S_{0,m}\over (1+m + \beta)}\cr
  &+\fr34 \sin^2\theta 
        \sum\limits_m
      {(2 + m)A_{1,m}S_{1,m} \over (2+m + \beta)}  \Bigr].\cr}
\eqno\new$$
As expected, this is a spherically aligned solution and all the
cross-terms of the velocity dispersion tensor vanish.

Now, the density of the scale-free power-law models can be expanded as:
\eqnam\powerdensityc
$$\eqalign{\rho = {1\over r^{2 + \beta}g^{2 + \beta/2}(\theta)}
\Bigl[& (2 - Q(1 + \beta)) g(\theta)\cr
+&(\beta+2)(Q-1)\sin^2\theta \Bigr].\cr}\eqno\new$$
The spherically aligned Jeans solution can be deduced from
(\wynseqa) by putting $H_2 =0$. We obtain:
\eqnam\powerkin
$$\eqalign{ \rho \langle v_r^2 \rangle =
& {(1+\beta)^{-1}\over r^{2 + 2\beta} g^{2+\beta} (\theta)}
                  \Bigl[ [\fr12 (2\!-\!Q(1+\beta))\!+
                H_1 q^2] g(\theta)\cr
& +(Q\!-\!1)(1+\beta)(\fr12
                   \!+\! H_1 q^2)\sin^2\theta\Bigr],\cr
           \rho \langle v_\theta^2 \rangle =
& {(1+\beta)^{-1}\over r^{2+2\beta} g^{2+ \beta} (\theta) }
\Bigl[ [\fr12 (2\!-\!Q(1+\beta)) -
\beta q^2H_1] g(\theta)\cr
& + \fr12(Q\!-\!1)(1+\beta)\sin^2 \theta\Bigr],\cr
            \rho \langle v_\phi^2 \rangle =
&{(1+\beta)^{-1}\over r^{2+2\beta} g^{2+\beta} (\theta) }
\Bigl[ [\fr12 (2\!-\!Q(1+\beta)) - \beta q^2H_1]
g(\theta)\cr
& + (Q\!-\!1)(1+\beta)(\fr32-2\beta q^2 H_1)\sin^2 \theta\Bigr].\cr}
\eqno\new$$
We must now choose the amplitudes $A_{0,m}$ and $A_{1,m}$
and the constant $C$ so that the densities (\dfdensity) and
(\powerdensityc) and the kinematics (\dfkin) and (\powerkin)
coincide. The calculation is straightforward and reveals that
$$C = -{6 \over 5} (Q-1).\eqno\new$$ 
This verifies that $C$ is small, at least for moderate flattenings
This justifies the first-order Taylor expansions used in our
analysis. For example, if the flattening is E3, then $q \sim 0.9$, and
so the next term in the Taylor expansion is $O(C^2)$ which is $\sim
8\%$.  The constraints concerning the unknown amplitudes $A_{0,m}$ and
$A_{1,m}$ are most compactly written by introducing the weights
\eqnam\defweights
$$\eqalign{
A_{0,m} S_{0,m} &= (2-Q(1+\beta)) W_{0,m}, \cr
A_{1,m} S_{1,m} &= (\beta +2)(Q-1) W_{1,m}. \cr}
\eqno\new$$
Then, there are three conditions on the $W_{0,m}$, namely
\eqnam\wzeroconds
$$\eqalign{&\sum\limits_m W_{0,m} =1,\cr
&\sum\limits_m {W_{0,m}\over 1 + m+ \beta}
      = {1\over 1 + \beta}
       + {2q^2 H_1\over (1+\beta)(2-Q(1+\beta))},\cr
&\sum\limits_m mW_{0,m} = {2\beta q^2 H_1 \over 3[2 - Q(1+\beta)]}.\cr}
\eqno\new$$
There are two conditions on the $W_{1,m}$, namely
\eqnam\woneconds
$$\eqalign{&\sum\limits_m W_{1,m} = 1 + {2\beta\over
                                         5(\beta +2)} q^2 H_1,\cr
           &\sum\limits_m {W_{1,m}\over 2 + m + \beta} = {1 + 2
q^2 H_1\over 2 + \beta}.\cr}\eqno\new$$
Provided we choose at least three of the weights $W_{0,m}$ and at
least two of the weights $W_{1,m}$, these linear equations
(\wzeroconds) and (\woneconds) can always be solved. Of course, it is
not guaranteed that the DFs formed by superposing the components are
positive definite. This is a hard matter to check analytically, but
easy to do on a finite grid with the computer. In practice, we did not
find positivity to be a real constraint here.  We have verified
numerically that some of the DFs for the model with $\beta = 0.18, q =
0.9$ are indeed positive definite. In particular, we examined the four
cases $H_1 = \pm 0.1$ and $\pm 0.3$, with the non-vanishing weights
chosen as $W_{0,0}, W_{0,10}, W_{0,20}, W_{1,0}$ and $W_{1,20}$. It
seems that, at least for moderate flattenings and anisotropies, many
of the three-integral DFs constructed according to this recipe are
everywhere positive in phase space.

The DFs (\approxdf) therefore seem to show that the spherically
aligned Jeans solutions ($H_2 =0$), which are marked by the bold
lines in Figs.~1 and~2, are physical. This conclusion certainly
holds good for tangentially anisotropic solutions with $H_1 \lta 
0$. When $H_1$ becomes large and positive, the stress tensor is
radially distended. This is caused by the presence of eccentric
radial orbits in the model, for which our partial integral is not
a good invariant. In this limit, it is unwise to put undue faith 
in our DF. In the spherical limit, $C=W_{1,m}=A_{1,m} =0$. This means
that the second sum in (\approxdf) vanishes. The partial integral
$I_3$ reduces to $L^2$ and so we recover the DFs given in Section 3 of
this paper, as we should. 

There is one further, somewhat special, limit that deserves
mention. When $\beta =2$, the scale-free power-law potential is not
useful for modelling galaxies, as it generates negative density
through Poisson's equation. Nonetheless, it is perfectly reasonable to
consider the properties of orbits in this supplied force field. The
case is exceptional because the Hamilton-Jacobi equation separates
(e.g., Landau \& Lifshitz 1960; Lynden-Bell 1962). So, in this limit,
there is an exact, global third integral of the form
\eqnam{\eddington}
$$I_{3,\rm g} = \fr12 \Bigl[ r^2 (v_\theta^2 + v_\phi^2) - {1\over g(\theta)}
\Bigr].\eqno\new$$
Now, our partial integral (\defpartint) does not reduce to this
global integral when $\beta =2$. This, however, is no problem. Any 
thin tube, which is a one-dimensional manifold, must possess conserved 
quantities additional to the globally defined integrals. The partial 
integral valid for thin tubes (\defpartint) need not therefore reduce 
to (\eddington). When $\beta =2$, we see that
$$I_3 = 2I_{3,\rm g}g(\theta) +1.\eqno\new$$
Any change in the partial integral $\Delta I_3$ therefore satisfies
$$\Delta I_3 = 2 I_{3,\rm g} \Delta g(\theta),\eqno\new$$
We see that this change must be small for the thin tubes, because
$I_{3,\rm g} \approx 0$ from (\intcond). Note, too, that it follows
from the form of (\eddington) that the velocity ellipsoid is
spherically aligned when $\beta =2$ (in fact, this is a special case
of Eddington's (1915) theorem). It is interesting that this somewhat
exceptional case seems to reinforce our arguments for the physical
nature of the spherically aligned Jeans solutions.

\eqnumber =1
\def\chaphead{\hbox{6.}}
\section{6 Conclusions} 

One way forward in the construction of three-integral galaxy models is
to exploit partial integrals. These are invariants of the motion
specific to particular orbital families. Phase space is therefore
broken up into patches, and each patch is covered by an appropriate
integral. This general picture is similar to that advocated by Binney
and co-workers (e.g., Binney 1994; Kaasalainen \& Binney 1994), who
fit phase space tori to different orbital families using different
mappings.  It also has some tradition in the theory of resonant orbits
(Gerhard \& Saha 1991; Dehnen \& Gerhard 1993).  The main result of
this paper is the identification of such a partial integral for the
thin and near-thin tubes in arbitrary axisymmetric scale-free
potentials. It is a generalisation of the total angular momentum. By
careful numerical orbit integration, the excellence of the partial
integral for the thin tubes in one family of models -- the scale-free
power-law models (E94) -- has been verified in detail. This
supplements earlier work on the scale-free logarithmic potential
(ZES). For all the orbits we have examined in flattened potentials,
this partial integral is {\it always} better conserved than the total
angular momentum

Approximate and partial integrals are needed in stellar dynamics for
the construction of distribution functions (DFs). For the particular
case of the scale-free power-law models, we carried out a detailed
analysis of how to build three-integral DFs. This yielded the 
following conclusions:

\medskip
\noindent
(1) If DFs depend on any approximate or partial integral, it is not
guaranteed that they satisfy the Jeans equations. To be sure of
building a galaxy model in which the motions of the stars balance the
force field, it is important to combine both the Jeans and Boltzmann
approaches. We have shown, both here and in ZES, that such combined
Jeans and Boltzmann solutions can be found in a simple way.

\medskip
\noindent
(2) The solution of the Jeans equations for arbitrary axisymmetric
scale-free models has been deduced. Jeans solutions in which the 
velocity ellipsoid is spherically aligned probably are physical, at least
for moderate flattenings and anisotropies. This holds because the 
partial integral is quadratic in the velocities and does not generate 
any cross-terms in the stress tensor. At present, it is not known 
whether any of the remaining Jeans solutions are physical. Even
at the level of the Jeans equations, it is evident that we cannot
construct very radially anisotropic models with strong cusps
($\beta >0$).

\medskip
\noindent
(3) DFs depending on the classical integrals $E$ and $L_z$ and the
partial integral for near-thin tubes $I_3$ can generate both the
density and the spherically-aligned second moments of the power-law
models. The scale-free power-law models are now a simple and useful
set of galaxy models which have both two-integral and three-integral
DFs readily available.

\medskip
\noindent
There are three pressing questions left unresolved by the work in ZES
and this paper. First, what is the appropriate partial integral to
choose for the radial orbits in scale-free models? Although the phase
space areas occupied by the thin and near-thin tubes are patched by
the partial integral deduced here, the areas occupied by the eccentric
radial orbits are not. The answer to this question is needed to build
radially anisotropic three-integral DFs. Second, although one sub-set
of the Jeans solutions displayed in Figs.~1 and~2 has been shown to be
physical by construction of suitable DFs, the status of the remaining
solutions is unclear. Is it possible that at least some -- perhaps
even all -- of the remaining Jeans solutions are physically realisable?
This is, of course, related to the first question, as the Jeans
solutions must be built by DFs depending on partial integrals
appropriate to other orbital families. Third, how does this work
generalise to fully triaxial scale--free models? Here, the solutions
of the Jeans equations are now explicitly available (Carollo, de Zeeuw
\& Evans 1996), but there are no known integrals other than
energy.

\section*{Acknowledgments}
NWE thanks the Royal Society for financial support. RMH is partially
supported by the Particle Physics and Astronomy Research Council. PTdZ
thanks Theoretical Physics, Oxford for their hospitality during a
working visit. We wish to thank Ortwin Gerhard for some perceptive
comments on scale-free models, as well as Massimo Stiavelli, Jos de
Bruijne and an anonymous referee for helpful criticism of the draft
manuscript.

\eqnumber =1
\def\chaphead{\hbox{A}}

\section{Appendix A: Jeans Solutions for Cored Models}

Here, we give the extension of solution (\wynseqa) to the stresses
associated with slightly more general axisymmetric densities in the
potentials of the cored power-law models (E94).

We consider the general density law:
$$\rho = {a_1 R^2 +c_1 z^2 +d_1 R_c^2 \over 
          (R_c^2 + R^2 + z^2/q^2)^{1+\gamma/2}},      \eqno\new$$
with $a_1$, $c_1$, $d_1$ and $\gamma$ free parameters. The potential
is
$$\Phi = \cases{ \fr12 \ln(R_c^2+R^2+z^2/q^2), & for $\beta=0$, \cr
                 \null & \null \cr
                -{\displaystyle 
                    R_c^\beta \over 
                   \displaystyle \beta (R_c^2+R^2+z^2/q^2)^{\beta/2}}, 
                                               & for $\beta \not=0$. \cr}
                                                                  \eqno\new$$
Here, $R_c$ is the core radius. When $R_c=0$ we recover the scale-free 
potentials (\genpot) with $g(\theta)$ given in (\powergtheta). The 
two-parameter solution of the Jeans equations is: 
\eqnam{\genjeans}
$$\eqalign{
\rho \langle v_R^2 \rangle 
  &= {1 \over (\alpha\!-\!1)} {A_1 R^2 + C_1 z^2 + D_1 R_c^2 
                         \over (R^2 + z^2/q^2 + R_c^2)^\alpha}, \cr
\rho \langle v_{R}v_{z}\rangle 
  &= {1 \over (\alpha\!-\!1)} {B_2 R z 
                         \over (R^2 + z^2/q^2 + R_c^2)^\alpha}, \cr
\rho \langle v_z^2 \rangle 
  &= {1 \over (\alpha\!-\!1)} {A_3 R^2 + C_3 z^2 + D_3 R_c^2 
                         \over (R^2 + z^2/q^2 + R_c^2)^\alpha}, \cr
\rho \langle v_\phi^2 \rangle 
  &= {1 \over (\alpha\!-\!1)} {A_4 R^2 + C_1 z^2 + D_1 R_c^2 
                         \over (R^2 + z^2/q^2 + R_c^2)^\alpha}. \cr}   
                                                                  \eqno\new$$
Here, $\alpha=(2+\gamma+\beta)/2$ and the relationships between
the coefficients and the $H_1$ and $H_2$ parameters are given in
Table~4.
\begintable{4}%
\caption{{\bf Table 4.} Coefficients for the Jeans solution (A3).}
\halign{#\hfil&\quad#\hfil\quad\cr
\noalign{\hrule}
\noalign{\vskip0.3truecm}
$A_{1}$         & $[(\alpha\!-\!1)\!+\!(2\!-\!\alpha)q^2]H_1 +\!
                  (\alpha\!-\!1)H_2$\cr
${}   $         & $\qquad+\!{\displaystyle (\alpha\!-\!1)a_1\!+\!q^2 c_1 \over 
\displaystyle 2\alpha}$\cr 
\noalign{\vskip0.2truecm}
$C_{1}$         & $(2\!-\!\alpha)H_1\! +\! {(\alpha\!-\!1)\over q^2}
H_2 \!+\! \fr12 c_1$ \cr
\noalign{\vskip0.2truecm}
$D_{1}$         & $[(\alpha\!-\!1)\!+\!(2\!-\!\alpha)q^2]H_1$\cr
${}   $         & $\qquad +\!(\alpha\!-\!1)H_2
        \!+\!{\displaystyle q^2c_1\!+\!(\alpha\!-\!1)d_1 \over 
              \displaystyle 2\alpha}$\cr
\noalign{\vskip0.2truecm}
$B_{2}$         & $(\alpha\!-\!1)H_1$ \cr
\noalign{\vskip0.2truecm}
$A_{3}$         & $(2\!-\!\alpha)q^2 H_1
         \!+\!{\displaystyle (\alpha\!-\!1)a_1 \!+\!q^2c_1 \over 
               \displaystyle 2\alpha}$\cr
\noalign{\vskip0.2truecm}
$C_{3}$ & $H_1 \!+\! \fr12 c_1$ \cr
\noalign{\vskip0.2truecm}
$A_{4}$ & $ (2\!-\!\alpha)[2\alpha\!-\!2\!+\!(3\!-\!2\alpha)q^2]H_1$ \cr
${}   $ & $\qquad+\!(\alpha\!-\!1)(3\!-\!2\alpha)H_2
        \! +\!{\displaystyle 3(\alpha\!-\!1)a_1\! +\!(3\!-\!2\alpha)q^2c_1 \over
\displaystyle 2\alpha}$\cr
\noalign{\vskip0.1truecm}
\noalign{\hrule}
}
\endtable

Self-consistent models have
$$\eqalign{
a_1    &= {1\over q^2} (1-\beta q^2), \cr
c_1    &= {1\over q^2} \bigl(2 - {1+\beta \over q^2}\bigr), \cr
d_1    &= {1\over q^2} + 2,\cr
\gamma &= \alpha = 2+\beta.\cr} \eqno\new$$
Upon substitution of these expressions, and taking the limit $R_c=0$,
we recover the results in Section 2.4 of the main text.

It is instructive to transform to spheroidal coordinates $(\lambda,
\phi, \nu)$, with $(\lambda, \nu)$ the two roots for $\tau$ of 
$R^2/(\tau-R_c^2) +z^2/(\tau-q^2R_c^2) =1$. The foci lie at 
$(R=0, z=\pm R_c \sqrt{1-q^2})$, and $\lambda$ and $\nu$ are given by 
$$\eqalign{
\lambda, \nu &= \fr12 [R^2+z^2+R_c^2(1+q^2)] \pm \cr 
             &\fr12 \sqrt{(R^2\!+\!z^2)^2 \!+\!2R_c^2(1\!-\!q^2)
                          (R^2\!-\!z^2)\!+\!R_c^4(1\!-\!q^2)^2}. \cr} 
                                                                  \eqno\new$$
The potential then simplifies to
$$\Phi = \cases{ \fr12\ln\lambda +\fr12\ln\nu +\ln qR_c, & for $\beta=0$, \cr
                 \null & \null \cr
                -{\displaystyle 1 \over \displaystyle \beta} 
                  \bigl( {\displaystyle R_c^4 q^2 \over 
                          \displaystyle \lambda \nu} \bigr)^{\beta/2}, 
                                               & for $\beta \not=0$. \cr}
                                                                  \eqno\new$$
The Jeans solutions (\genjeans) can be rewritten to give $\rho
\langle v_\lambda^2 \rangle$, $\rho \langle v_\lambda v_\nu \rangle$,
and $\rho \langle v_\nu^2 \rangle$ by means of the relations between
$(v_R, v_z)$ and $(v_\lambda, v_\nu)$ given in, e.g., Dejonghe \& de
Zeeuw (1988). In particular, we find
$$(\lambda-\nu) \rho \langle v_\lambda v_\nu \rangle = 
  - {H_2 Rz \over (R_c^2 + R^2 + z^2/q^2)^{\alpha-1}}.        \eqno\new$$
This allows us to consider three limiting cases at once. 

When $q=1$ the potential -- but not necessarily the density -- is
spherical. The foci of the spheroidal coordinates now coincide with
the origin, and hence $(\lambda, \phi, \nu)$ reduce to ordinary
spherical coordinates, with $v_\lambda=v_r$ and $v_\nu =
-v_\theta$. It follows that $\langle v_r v_\theta \rangle \propto
H_2$. But in a spherical potential this cross term in the stress
tensor must vanish by the symmetries of the individual stellar orbits,
so that the only Jeans solutions with physical DFs are those with
$H_2=0$. This result is valid for arbitrary $\beta$ and $R_c$.

When $\beta=2$ the potential is of St\"ackel form in the coordinates
$(\lambda, \phi, \nu)$, for arbitrary $q$ and $R_c$ (de Zeeuw \&
Lynden--Bell 1985). It then admits three exact integrals of motion
that are quadratic in the velocity components, and any DF will
therefore give $\langle v_\lambda v_\nu \rangle \equiv 0$.  In this
limit we must therefore also restrict the Jeans solutions (\genjeans)
to those with $H_2=0$, irrespective of the values of $q$ and $R_c$.

When $R_c=0$ the spheroidal coordinates again reduce to sphericals,
but the potential is flattened unless $q=1$. In this case the model
potential is scale-free, but there is no restriction on $H_2$, unless
$q=1$ or $\beta=2$.

The above shows that if we consider our Jeans solutions as functions
of the parameters $q$ and $\beta$, then along the two boundaries $q=1$
and $\beta=2$ of the parameter-space they must reduce to a
one-parameter family, with $H_2=0$. There appears to be no physical
reason for a similar restriction in the entire $(q, \beta)$-plane.

Finally, we remark that the line-of-sight projected second moments 
associated with the solutions (\genjeans) are all of the general 
form (3.7) of Evans \& de Zeeuw (1994), where the coefficients 
$a_{ij}, b_{ij}, c_{ij}$ and $d_{ij}$ can be worked out explicitly.

\eqnumber =1
\def\chaphead{\hbox{B}}

\section{Appendix B: Constraints on the Positivity of the Principal
Stresses}

In this appendix, we write out explicitly the constraint that the 
two-parameter Jeans solutions discussed in Sections 2.4 and 2.5 have
positive principal components. The areas inside the following four
curves must be deleted from Figs.~1 and~2.
$$H_2 = {(1\!-\!Q)(Q^2\!-\!Q\!+\!\beta(Q^2\!+\!Q\!+\!2H_1))^2 \over
2Q(1\!+\!\beta)(2QH_1\!-\!Q(Q\!-\!1)^2\!+\!\beta(2H_1\!+\!Q\!-\!Q^3))}.$$
$$\eqalign{&H_2={(1\!-\!Q)(Q^2\!-\!2Q\!-\!H_1\!+\!\beta Q^2\!+\!\beta H_1)
\over Q(1\!+\!\beta)} \times \cr
&{Q^2\!-\!Q\!+\!2\beta H_1\!+\!\beta Q(1\!+\!Q) \over Q(5Q\!-\!2Q^2\!+\!4H_1\!-\!3)\!-\!2H_1\!
+\!\beta (2H_1\!+\!Q\!+\!Q^2\!-\!2Q^3)}.\cr}$$
$$\eqalign{&H_2 ={(1\!-\!Q)(Q\!+\!H_1Q\!+\!\beta(H_1Q\!-\!2H_1\!-\!Q)\over
Q(1\!+\!\beta)} \times \cr
&{\beta(Q^2\!+\!Q\!+\!2H_1) -Q(1\!-\!Q)\over
Q(2\!-\!2H_1\!-\!3Q\!+\!Q^2)\!
+\!\beta(Q^3\!+\!Q^2\!+\!2H_1Q\!-\!2Q\!-\!4H_1)}.\cr}$$
$$\eqalign{&H_2 = {H_1(1\!-\!Q)\over Q(1\!+\!\beta)(2H_1\!+\!Q\!-\!Q^2)^2}\times\cr
&\Bigl[ (Q(Q^3\!-\!2Q^2\!-\!Q\!-\!2H_1(1\!+\!Q)\!+
\!\beta(Q^2\!+\!Q^4 \!+\!4H_1^2\!+\!4QH_1))
\cr
&\pm 2Q[(2H_1\!+\!Q)(Q\!-\!2\beta H_1\!-\!\beta Q)
(2H_1\!+\!2Q\!-\!Q^2\!-\!\beta Q^2)]^{\fr12}\Bigr].\cr}$$
These lengthy expressions are readily evaluated numerically. A quick
check on their accuracy is provided by taking the limit $\beta =0$,
when they reduce to equations (2.38) of ZES.

\eqnumber =1
\def\chaphead{\hbox{C}}

\section{Appendix C: Triaxial Components with Triaxial Kinematics}

Almost certainly, the potentials of galaxy haloes are not exactly
spherical. But, one of the very attractive reasons for using spherical
potentials is the ready availability of four functionally-independent
integrals of the motion $E, L_x, L_y$ and $L_z$. As realised clearly
by White (1985) and Arnold (1992), this is useful for building 
axisymmetric tracer populations with triaxial kinematics -- as is 
warranted by the shape of the local velocity ellipsoid in the Milky
Way, for example. The same methods can also be exploited to give 
{\it triaxial} tracer populations with triaxial kinematics! (c.f., 
Mathieu, Dejonghe \& Hui 1996).

Together with White (1985), the work in Section 3 shows that 
the DFs 
$$f_{m,n} (E,L^2,L_z^2) = \cases{\eta_{m, n} L^{2m} L_z^{2n}
|E|^{\alpha},& if $\beta \neq 0$,\cr \eta_{m, n} L^{2m} L_z^{2n} 
\exp(-E),& if $\beta = 0$,\cr}\eqno\new$$
generate the density law
$$\rho =  r^{-\gamma} \sin^{2n}\theta,\eqno\new$$
in the spherical scale-free power-law potentials
$$\Phi = \cases{ -{\displaystyle 1\over \displaystyle \beta r^\beta},& 
                   if $\beta \neq 0$,\cr
                   \null&\null\cr
                   \log r,& if $\beta =0$.\cr}\eqno\new$$
The constant $\eta_{m,n}$ is available as (\defetamn) when $\beta
\neq 0$. For the logarithmic case ($\beta =0$), ZES show that
$$\eta_{m,n} = {(m\!+\!n\!+\!\fr{\gamma}{2})^{m+n+{3\over2}}
               \Gamma(n\!+\!1) \over
               \pi \Gamma(n\!+\!\fr12) \Gamma(n\!+\!m\!+\!1)}.
\eqno\new$$

But, there is nothing special about the $z$-axis, and no reason for
the $z$-component of angular momentum to play a distinguished
role! It is straightforward to show that the DFs
$$g_{m,n} (E,L^2,L_x^2) = \cases{\eta_{m, n} L^{2m} L_x^{2n}
|E|^{\alpha},& if $\beta \neq 0$,\cr \eta_{m, n} L^{2m} L_x^{2n}
\exp(-E),& if $\beta = 0$,\cr}\eqno\new$$
give the stellar density
$$\rho = r^{-\gamma} \Bigl[ \sin^2 \theta \sin^2 \phi + \cos^2\theta
\Bigr]^n,\eqno\new$$ 
and the kinematics
\eqnam{\kinematappb}
$$\eqalign{
&\langle v_r^2 \rangle = {v_{\rm circ}^2 \over 2m + 2n + \bandg }, \cr
&\langle v_\theta^2 \rangle = {m + n + 1 \over n+1}
{ (2n+1)\sin^2\phi + \cos^2\theta\cos^2\phi 
\over \sin^2\phi + \cos^2\theta \cos^2\phi} \langle v_r^2 \rangle, \cr
&\langle v_\phi^2 \rangle = {m + n + 1 \over n+1}
{ \sin^2\phi + (2n+1)\cos^2\theta\cos^2\phi \over \sin^2\phi + 
\cos^2\theta \cos^2\phi} \langle v_r^2 \rangle, \cr
&\langle v_\theta v_\phi \rangle = {n (m+n+1) \over n+1}
{\sin\phi \cos\phi \cos\theta \over  \sin^2\phi + \cos^2\theta 
\cos^2\phi}\langle v_r^2 \rangle .\cr}\eqno\new$$
Equally, the DFs
$$h_{m,n} (E,L^2,L_y^2) = \cases{\eta_{m, n} L^{2m} L_y^{2n}
|E|^{\alpha},& if $\beta \neq 0$,\cr \eta_{m, n} L^{2m} L_y^{2n}
\exp(-E),& if $\beta = 0$,\cr}\eqno\new$$
correspond to
$$\rho = r^{-\gamma} \Bigl[ \sin^2 \theta \cos^2 \phi + \cos^2\theta
\Bigr]^n,\eqno\new$$
with the kinematics (\kinematappb) on making the modifications
$\cos \phi \rightarrow \sin \phi$ and $\sin \phi \rightarrow
-\cos \phi$.
Using the method of superposition of components, a very general DF
corresponding to a triaxial halo with a triaxial velocity ellipsoid 
is
$$\eqalign{ f(E, L_x^2, L_y^2, L_z^2 ) =& \sum_{m,n} A_{m,n}
f_{m,n}(E,L^2,L_z^2) \cr
+& \sum_{m,n} B_{m,n}g_{m,n}(E,L^2,L_x^2) \cr
+& \sum_{m,n} C_{m,n}h_{m,n} (E,L^2,L_y^2).\cr}\eqno \new$$
The components can find a ready application for fitting the data 
on the kinematics of the stellar halo of the Milky Way.

\bye